\documentclass[prb,twocolumn,amsmath,english,amssymb,superscriptaddress]{revtex4-1}
\usepackage{epsfig, graphicx,graphics,amsmath,amssymb,float}
\usepackage[T1]{fontenc}
\usepackage{bm}
\usepackage{amsmath}
\usepackage{amssymb}
\usepackage{graphicx}
\usepackage{esint}
\usepackage[unicode=true, bookmarks=false, breaklinks=false,pdfborder={0 0 1},backref=false,colorlinks=false] {hyperref}
\usepackage{amscd}
\usepackage{psfrag}
\usepackage{bbm} 
\usepackage{babel}
\usepackage{wasysym}
\usepackage{mathrsfs}
\usepackage{color}

\makeatletter
\usepackage{bbold}
\usepackage{soul}
\usepackage{epsfig}
\usepackage{verbatim}
\usepackage{array}
\usepackage{braket} 
\newcommand{\be}{\begin{equation}}
\newcommand{\ee}{\end{equation}}
\newcommand{\bea}{\begin{eqnarray}}
\newcommand{\eea}{\end{eqnarray}}
\newcommand{\mc}{\mathcal}
\newcommand{\mb}{\mathbf}

\begin{document}

\title{Scanning tunneling spectroscopy of Majorana zero modes in a Kitaev spin liquid}

\author{Tim Bauer}
\affiliation{Institut f\"ur Theoretische Physik,
Heinrich-Heine-Universit\"at, D-40225  D\"usseldorf, Germany}
\affiliation{International Institute of Physics and Departamento de F\'isica Te\'orica
e Experimental, Universidade Federal do Rio Grande do Norte, 
Natal, RN, 59078-970, Brazil}
\author{Lucas R. D. Freitas} 
\affiliation{International Institute of Physics and Departamento de F\'isica Te\'orica
e Experimental, Universidade Federal do Rio Grande do Norte, 
Natal, RN, 59078-970, Brazil}
\affiliation{Institut f\"ur Theoretische Physik,
Heinrich-Heine-Universit\"at, D-40225  D\"usseldorf, Germany}
\author{Rodrigo G. Pereira}
\affiliation{International Institute of Physics and Departamento de F\'isica Te\'orica
e Experimental, Universidade Federal do Rio Grande do Norte, 
Natal, RN, 59078-970, Brazil}
\author{Reinhold Egger}
\affiliation{Institut f\"ur Theoretische Physik,
Heinrich-Heine-Universit\"at, D-40225  D\"usseldorf, Germany}
\begin{abstract}
We describe scanning tunneling spectroscopic signatures of Majorana zero modes (MZMs) in Kitaev
spin liquids. The tunnel conductance is determined by the dynamical spin correlations of the spin liquid, 
which we compute exactly, and by spin-anisotropic cotunneling form factors.
Near a $\mathbb{Z}_2$ vortex, the tunnel conductance has a staircase voltage dependence,
where conductance steps arise from MZMs and (at higher voltages) from additional vortex configurations.
By scanning the probe tip position, one can detect the vortex locations. 
Our analysis suggests that topological magnon bound states near defects or magnetic impurities generate
 spectroscopic signatures that are
qualitatively different from those of MZMs.
\end{abstract}
\maketitle

\section{Introduction}\label{sec1}

Presently a major goal in condensed matter physics is 
to realize, detect, and manipulate 
topologically ordered phases of frustrated quantum magnets, commonly referred to as
quantum spin liquids (QSLs).
A famous exactly solvable paradigm is given by Kitaev's 
two-dimensional (2D) honeycomb lattice spin model with bond-dependent anisotropic exchange which,
in a magnetic field,  describes a gapped non-Abelian chiral QSL \cite{Kitaev2006}.  
Emergent excitations of the Kitaev spin liquid include MZMs bound to $\mathbb{Z}_2$ vortices (``visons''), which are Ising anyons of interest for quantum information processing,
as well as gapped bulk fermions and a chiral Majorana edge mode at the boundary.  
Being excitations of an insulating magnet, they are electrically neutral.
Sizable Kitaev couplings are expected \cite{Jackeli2009} and have been reported in various material platforms 
for Mott insulators with strong spin-orbit coupling, e.g., in iridate compounds 
or in $\alpha$-RuCl$_3$, where the smallness of interlayer couplings justifies the use of 2D models.
For recent reviews, see Refs.~\cite{Savary2017,Zhou2017,Wen2017,Winter2017,Hermanns2018,Knolle2019,Takagi2019,Motome2020,Broholm2020,Trebst2022}.  Despite the impressive experimental progress achieved over the past 
decade, however, no consensus has emerged  whether $\alpha$-RuCl$_3$ or any other known material harbors a QSL. 
In particular, the half-quantized thermal Hall conductivity due to the chiral Majorana edge mode
reported in Refs.~\cite{Kasahara2018,Yokoi2021,Bruin2022} has not been found 
in other experiments \cite{Nagler2021,Czajka2022}. In fact, some spin-liquid predictions can be mimicked by topological magnons in a polarized phase \cite{Kim1,Kim2,Wulferding2020}.

\begin{figure}
\begin{center}
\includegraphics[width= \columnwidth]{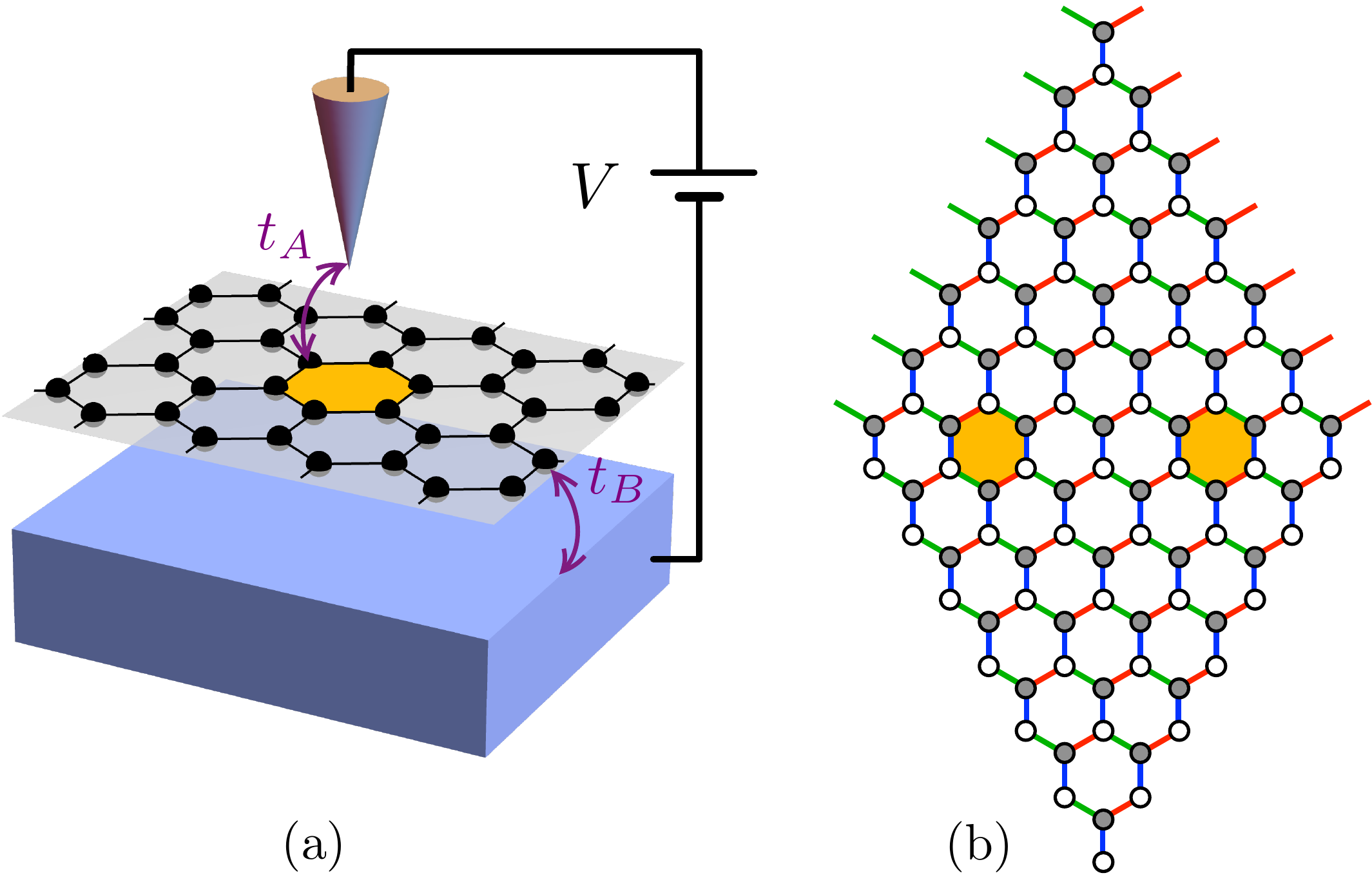}
\end{center}
\caption{(a) Schematic STS setup. Tunnel couplings $t_A$ and $t_B$ connect the QSL layer to the tip and the substrate, respectively. The differential conductance $G(V)=\frac{dI}{dV}$ follows by measuring the tunnel current $I$ from
tip to substrate as function of the applied voltage $V$. (b) Finite 2D Kitaev honeycomb lattice with $L\times L$ unit cells and periodic boundary conditions, shown for $L=7$ and a configuration $\cal G$ 
with two $\mathbb{Z}_2$ vortices (shaded).
Full and open circles represent the two sublattices. Nearest-neighbor bonds $\langle jl\rangle_\alpha$
of type $\alpha\in\{x,y,z\}$ are distinguished by different colors.}
\label{fig1}
\end{figure}

We here show that characteristic signatures of Ising anyons should be seen in  
scanning tunneling spectroscopy (STS) experiments \cite{STMreview} on a 2D Kitaev layer \cite{Ziatdinov2016,Weber2016,Du2018,Ruan2021}
by scanning the probe-tip position in the vicinity of an isolated $\mathbb{Z}_2$ vortex 
(located far away from all other vortices and from the sample boundary) and/or by changing the applied voltage, see Fig.~\ref{fig1}. 
Below we will also compare our results to an alternative scenario 
with topological magnon bound states near defects or magnetic impurities, 
which could also cause low-energy features in the STS tunnel conductance.  
Such a comparison is important as evidenced by the corresponding topological superconductor case \cite{Alicea2012}, where the tunnel conductance has a zero-bias anomaly with quantized peak conductance $2e^2/h$ due to MZM-mediated resonant Andreev reflection  \cite{Sengupta2001,Law2009,Flensberg2010,Zazunov2016}. 
STS experiments have found such zero-bias anomalies near vortex cores in various superconducting materials
and attributed them to MZMs \cite{STMreview,Machida2018,Liu2018,Kong2019,Zhu2020}.
A major obstacle to this interpretation is that very similar conductance peaks can be caused by conventional
disorder-induced Andreev bound states \cite{Prada2020}.  
However, the magnetic QSL case is rather different and warrants a separate investigation. 
The absence of a Cooper pair condensate implies that the charge of an electron 
(tunneling in from the tip via the MZM) is much harder to accomodate. 
For the pure Kitaev model, the infinite charge gap implies a vanishing tunnel conductance, $G(V)=0$.

To obtain a finite $G(V)$, we start from the Hubbard-Kanamori model for Kitaev materials 
\cite{Jackeli2009,Rau2014,Rau2016,Winter2016,Pereira2020}. Adding a
tunneling Hamiltonian for the QSL couplings to tip and substrate, see Fig.~\ref{fig1}(a), and projecting to states with energy below the charge gap, we obtain $H=H_K+H_{\rm cot}$, where $H_K$
describes the Kitaev model \cite{Jackeli2009} and the cotunneling Hamiltonian $H_{\rm cot}$ 
encodes tip-substrate electron transfer
due to virtual excursions to high-energy intermediate states~\cite{Fernandez2009,Fransson2010,Delgado2011}. 
We compute $H_{\rm cot}$ for arbitrary tip position and find that it is anisotropic in spin space.
One then obtains $G(V)$ from the dynamical spin correlations of the QSL \cite{Feldmeier2020,Koenig2020,Carrega2020,Chen2020,Udagawa2021},
which can be computed exactly  \cite{Baskaran2007,Pedrocchi2011,Knolle2014,Zschocke2015,Song2016,Hassler2019}.
However, in the presence of $\mathbb{Z}_2$ vortices, we encounter a technical challenge 
described and resolved below.  

As a function of voltage, we predict a characteristic sequence of conductance steps linked to MZMs. 
By scanning the tip location at fixed voltage, one can locate MZMs in real space and obtain information about the
vortex configurations contributing to the conductance.
It stands to reason that experimental tests of our theory will help in identifying QSLs. 
(For other proposals aimed at the electric detection of QSLs, see Refs.~\cite{Aasen2020,Pereira2020,Yamada2021,Chari2021,Banerjee2022}.)
Our study of an alternative topological magnon scenario suggests 
that MZM signatures obtained by STS on a Kitaev layer are easier to distinguish 
from other mechanisms than in the superconducting case.

The structure of the remainder of this article is as follows.  In Sec.~\ref{sec2}, we derive the low-energy theory used for calculating the differential conductance,
where technical details have been relegated to App.~\ref{appA}.
We then show in Sec.~\ref{sec3} how to compute the conductance in terms of an exact evaluation of dynamical spin-spin correlation functions of the Kitaev layer. 
Our results for the conductance profile are shown in Sec.~\ref{sec4}. In Sec.~\ref{sec5}, we then address a complementary topological magnon scenario.
Finally, we offer concluding remarks in Sec.~\ref{sec6}.

\section{Effective low-energy theory}\label{sec2}

We consider the setup in Fig.~\ref{fig1}(a), where a scanning probe tip at position $\mb r=(x,y,d)$ is tunnel-coupled to a 2D Kitaev layer at vertical distance $d$.  
The layer is also coupled to a metallic substrate. Throughout, we assume weak and spin-independent tunnel amplitudes. Due to the charge gap in the magnetic layer, electron transport
at subgap voltages $V$, applied between the tip (with conduction electron creation 
operator $\Psi^\dagger_{A\tau}(\mb r)$ for spin projection $\tau=\uparrow,\downarrow$)
and the substrate (with $\Psi^\dagger_{B\tau}(\mb R_j)$ below lattice site $\mb R_j$),
can only take place via cotunneling  \cite{Fransson2010,Feldmeier2020,Elste2007}.   
We use the Hubbard-Kanamori model for strongly correlated $d^5$ electrons in  $\alpha$-RuCl$_3$ or related materials 
\cite{Jackeli2009,Rau2014,Rau2016,Pereira2020}, where on-site correlations are 
captured by a large Coulomb energy $U$ and a Hund coupling $J_H$. 
Including a tunneling Hamiltonian for the contacts to tip and substrate, the projection to  energies below the charge gap $\sim U$ can be performed by a 
canonical transformation~\cite{Jackeli2009,Pereira2020}. 
We show this calculation in some detail in App.~\ref{appA}.

The low-energy theory is described by spin-$1/2$ operators, $\mb S_j=\frac12 \boldsymbol\sigma_j$, in the QSL layer, where $H=H_K+H_{\rm cot}$ includes the 
 Kitaev model \cite{Kitaev2006,Jackeli2009}
\begin{equation}\label{Kitaev}
    H_K= -K \sum_{\langle jl\rangle_\alpha} \sigma_j^\alpha \sigma_l^\alpha 
    -\kappa \sum_{\langle jk\rangle_\alpha,\langle kl\rangle_\beta} \sigma_j^\alpha \sigma_k^\gamma \sigma_l^\beta,
\end{equation}
with $\langle jl\rangle_\alpha$ denoting a nearest-neighbor bond of type $\alpha\in \{x,y,z\}$, see Fig.~\ref{fig1}(b).
The term $\propto \kappa$ describes a magnetic field \cite{Kitaev2006,Song2016},
where $(\alpha\beta\gamma)$ is a cyclic permutation of $(xyz)$ and the sum runs over triangles $(jkl)$ 
with two adjacent nearest-neighbor bonds. We measure lengths in units of the lattice 
spacing $a_0$, where $a_0\approx 5.9$\AA~for $\alpha$-RuCl$_3$  \cite{Kaib2021}.  
The projection scheme yields a ferromagnetic (positive)  Kitaev coupling $K\propto J_H$ \cite{Jackeli2009},   where experimental analysis gives $K\approx 5$~meV for $\alpha$-RuCl$_3$ \cite{Winter2017b}.  Theoretical estimates for $K$ in different Kitaev materials have been reported in Refs.~\cite{Winter2016,Sugita2020,Yang2022,Hou2017,Katukuri2014,Yamaji2014,Katukuri2015}, see   Table \ref{table1}.

\begin{table}[t]
\begin{center}
\begin{tabular}{c|c|c} \hline\hline
Material & $K$~(meV) & Method \\ \hline \hline 
$\alpha$-RuCl$_3$ & 5.0 & experimental analysis \cite{Winter2017b} \\ 
& 6.7 & exact diagonalization \cite{Winter2016}\\
& 8.0-8.25 & ab initio \cite{Sugita2020,Yang2022}\\
& 10.6 & density functional theory \cite{Hou2017}\\ 
\hline
Na$_2$IrO$_3$  & 16.8 & exact diagonalization \cite{Winter2016}  \\ 
& 16.9 & quantum chemistry methods \cite{Katukuri2014} \\
& 29.4 & perturbation theory \cite{Yamaji2014} \\ \hline
$\alpha$-Li$_2$IrO$_3$ & 6.3-9.8 & exact diagonalization \cite{Winter2016}\\ \hline
Li$_2$RhO$_3$& 2.9-11.7 & quantum chemistry methods \cite{Katukuri2015} \\ \hline\hline
\end{tabular}
\caption{Kitaev couplings reported from different methods for several materials. \label{table1}}
\end{center}
\end{table}

Similarly, summing over all lattice sites, the cotunneling Hamiltonian follows as
\begin{eqnarray}\label{Hcot}
    H_{\rm cot} &=&  \sum_j  \Psi^\dagger_{A}(\mb r) 
    \Bigl[ T_0(\mb r-\mb R_j) \mathbbm{1}_j + \\ &+& \mb T(\mb r-\mb R_j) \cdot \boldsymbol \sigma_j \Bigr]
    \Psi^{}_{B} (\mb R_j) + {\rm h.c.},\nonumber
\end{eqnarray}
where $\boldsymbol \sigma_j$ and $\mathbbm{1}_j$ act in Kitaev spin space. 
The $2\times 2$ matrices $T_0$ and $T^\alpha$, with  
$\mb T=(T^x,T^y,T^z)$, act in conduction electron spin space.  
All $T$ matrix elements scale $\propto t_A t_B/U$, 
with real-valued tunnel couplings $t_A$ ($t_B$) from tip (substrate) to a given site.
We assume a constant substrate coupling $t_B$. 
The tip couplings depend on the overlap between the spherically symmetric tip wave function 
and the respective $t_{2g}$-orbital (labeled by $\alpha=x,y,z$) for the $d^5$ electrons.
With an energy scale $t_0$ and a tunneling length $l_0\lesssim a_0$, 
we write \cite{Fransson2010,Feldmeier2020}
\be\label{taj}
t_{A\alpha}(\mb r,\mb R_j) = t_0 e^{-|\mb r\pm \mb v_\alpha -\mb R_j|/l_0},
\ee
with the overall  coupling $t_A\equiv \sqrt{t_{Ax}^2+t_{Ay}^2+t_{Az}^2}$.
The vectors $\mb v_\alpha$ with $|\mb v_\alpha|\approx 0.1 a_0$ encode the orbital overlaps, 
where the $\pm$ signs in Eq.~\eqref{taj} label the sublattice type of site $\mb R_j$, see App.~\ref{appA}.
The exponential scaling in Eq.~\eqref{taj} implies that only a few sites near 
the tip location $\mb r$ contribute.  Analytical but lengthy expressions 
for $T_0$ and $\mb T$ are given in App.~\ref{appA}.
 
Simpler results emerge by approximating $\mb v_\alpha=0$, which gives 
exact results for a tip located on top of a lattice site and otherwise causes deviations $\sim 10$\% in the tunnel couplings. (For the figures shown below, we
have used the full expressions.) We then obtain
\bea\label{cotfin}
&& H_{\rm cot}=  \sum_j \frac{t_A(\mb r-\mb R_j) t_B}{U} \Psi^\dagger_{A}(\mb r) 
\Bigl[\eta_0 +\eta_1 \boldsymbol \tau\cdot \boldsymbol \sigma_j 
 +\\ &&\quad +\,\eta_2(\tau^x+\tau^y+\tau^z)
 (\sigma^x+\sigma^y+\sigma^z)_j   \Bigr]\Psi^{\phantom\dagger}_{B}(\mb R_j)
 + {\rm h.c.}\nonumber
\eea
with $\frac{J_H}{U}$-dependent numbers $\eta_{j}\sim {\cal O}(1)$, see App.~\ref{appA}.
The SU$(2)$ spin rotation symmetry assumed in Refs.~\cite{Feldmeier2020,Koenig2020,Carrega2020} is in fact
lowered to a $\mathbb Z_3$ symmetry around the [111] axis.

\section{Differential conductance}\label{sec3}

At this point, it is straightforward to compute the differential conductance, $G(V)=\frac{dI}{dV}$, from Fermi's golden 
rule \cite{Feldmeier2020,Koenig2020,Elste2007}.  In the zero-temperature limit, we find
\be
G(V) = \sum_{jl,\alpha\beta}  C_{jl}^{\alpha\beta}({\bf r})
\int_0^{eV} d\omega  S^{\alpha\beta}_{jl}(\omega)
=\frac{e^2}{\hbar}\int_0^{eV} d\omega  S_G(\omega),\label{tunnelcond}
\ee
with the dynamical spin correlation function of the QSL, 
\begin{equation}\label{dyncor}
S_{jl}^{\alpha\beta}(\omega) = \int \frac{dt}{2\pi} e^{i\omega t}
\langle \Phi|\sigma_j^\alpha(t) \sigma_l^\beta(0)|\Phi\rangle.
\end{equation}  
The second step in Eq.~\eqref{tunnelcond} defines the averaged dynamical spin correlator $S_G(\omega)$, 
which follows by weighting $S_{jl}^{\alpha\beta}(\omega)$ with its form factor,
\be\label{cab}
C^{\alpha\beta}_{jl}(\mb r) = \frac{2e^2 d_A d_B}{\hbar} 
{\rm Tr}\left[ T^\alpha (\mb r-\mb R_j) T^\beta(\mb r-\mb R_l) \right],
\ee
with the tip (substrate) density of states $d_{A}$  ($d_B$) and a trace over conduction electron spin space.
Note that $\frac{dG}{dV}\propto S_G(V).$ The term $\propto T_0$ in Eq.~\eqref{Hcot}  generates a voltage-independent background 
(including a mixing term of $T_0$ and $\mb T$) not contained in Eq.~\eqref{tunnelcond}.  However, this term is insensitive 
to $\mathbb{Z}_2$ vortices and can be disentangled from Eq.~\eqref{tunnelcond}. 

The correlation function \eqref{dyncor} can be computed exactly for $H_K$ by means of a Majorana representation of the spin degrees of freedom
\cite{Baskaran2007,Pedrocchi2011,Knolle2014,Zschocke2015}. 
By writing $\sigma_j^\alpha=ic_jc_j^\alpha$ in terms of Majorana fermions with a local parity
constraint, $D_j=c_jc_j^xc_j^yc_j^z=+1$, one obtains  an exactly solvable noninteracting Hamiltonian for ``matter'' Majorana fermions, $\{ c_j\}$,
which move in a conserved $\mathbb{Z}_2$ gauge field $u_{\langle jl\rangle_\alpha}=ic_j^\alpha c_l^\alpha=\pm 1$ \cite{Kitaev2006},
\be\label{HMajorana}
H_K=iK\sum_{\langle jl\rangle_\alpha} u_{\langle jl\rangle_\alpha}c_jc_l-i\kappa 
\sum_{\langle jk\rangle_\alpha, \langle kl\rangle_\beta}u_{\langle jk\rangle_\alpha}u_{\langle kl\rangle_\beta}c_jc_l.
\ee 
All eigenstates of $H_K$ can be written as a projected tensor product of a matter fermion state, $|\varphi({\cal G})\rangle$, for given static gauge field configuration $|\cal G\rangle$,
\begin{equation}
|\Phi\rangle={\cal P}|{\cal G}\rangle |\varphi({\cal G})\rangle,
\end{equation}
with $H_K|\Phi\rangle=E_\Phi |\Phi\rangle=E_{\varphi({\cal G})}|\Phi\rangle$, where
the projection ${\cal P}=\prod_j\frac{1+D_j}{2}$ projects onto the physical subspace.
Defining gauge-invariant plaquette operators,
\begin{equation}
    W_p=\prod_{\langle jl\rangle_\alpha\in p} u_{\langle jl\rangle_\alpha}=\pm 1,
\end{equation} 
the ground state has $W_p=+1$ for all hexagonal plaquettes $p$ \cite{Kitaev2006}. 
Plaquettes with $W_p=-1$ then define $\mathbbm{Z}_2$ vortices,  which are expected near vacancies or magnetic impurities \cite{Dhochak2010,Willans2011,Vojta2016}
and harbor MZMs.
In order to study the case shown in Fig.~\ref{fig1}(a),
we  will then consider $|\Phi\rangle$ as the matter ground state~$|\varphi_0({\cal G})\rangle$ 
for a gauge configuration $\cal G$ with two well-separated $\mathbb{Z}_2$ vortices.  
We note that ${\cal G}$ can be constructed from a zero-vortex configuration ${\cal G}_0$ 
(with all bond variables $u_{\langle jl\rangle_\alpha}=+1$ for $j$ in sublattice ${\cal A}$ and 
$l$ in sublattice ${\cal B}$) by reversing the bond variables along an arbitrary string connecting both vortices.

For explicit calculations, we consider a finite honeycomb lattice with
$L\times L$ unit cells and periodic boundary conditions. 
The $2N=2L^2$ matter Majoranas are written as $c_j=c_\lambda(m,n)$, 
where $\lambda\in ({\cal A},{\cal B})$ labels the sublattice and $m,n=1,\ldots,L$ the unit cell 
at $\mb R_j=m\hat{\mb e}_1+n\hat{\mb e}_2$, with the primitive
lattice vectors  
$\hat{\mb e}_1=\frac12\hat{\mb x}+\frac{\sqrt3}{2}\hat{\mb y}$ and  $\hat{\mb e}_2=-\frac12\hat{\mb x}+\frac{\sqrt3}2\hat{\mb y}$.
We next define the $2N$-dimensional Majorana vector
$c=\left(c_{\cal A},c_{\cal B}\right)^T$, with the ordering convention
$c_\lambda=\left(c_\lambda(1,1),\ldots c_\lambda(L,1), c_\lambda(1,2),\ldots,
c_\lambda(L,L)\right)^T$,
and a complex fermion for each unit cell,
$f(m,n)=\frac12[c_{\cal A}(m,n)-ic_{\cal B}(m,n)]$. With an $N$-dimensional vector $f$ 
formed in analogy to $c_\lambda$, the linear transformation between both representations is given 
by 
\be\label{ctrans}
c =T\left(\begin{array}{c}
f\\ f^\dagger
\end{array}\right),\quad T=\left(\begin{array}{cc}
 {\mathbb 1}_N&{\mathbb 1}_N \\
i {\mathbb 1}_N& -i{\mathbb 1}_N\end{array}\right),
\ee
with the $N\times N$ identity ${\mathbb 1}_N$ and $T^{-1}=\frac12 T^\dagger$.
The projection ${\cal P}$ here implies a parity constraint
 for the total number $N_f$ of $f$ fermions and the total number 
$N_\chi$ of bond fermions $\chi^{}_{\langle jl\rangle_\alpha}=\frac12\left(c^\alpha_j-ic^\alpha_l\right)$
 \cite{Pedrocchi2011,Knolle2014,Zschocke2015}, 
\be \label{parityconstraint}
(-1)^{N_f+N_\chi}=1 ,
\ee
where we assume a vanishing boundary condition twist parameter in Ref.~\cite{Zschocke2015}. 
We note that $N_\chi$ is uniquely determined by the bond variables $\{ u_{\langle jl\rangle_\alpha} \}$ defining
the gauge configuration $\cal G$. Using the $f$ fermions, we obtain
\be\label{BdG}
H_K =\frac12 (f^\dagger \; f)\;T^\dagger \left(\begin{array}{cc}\mc H^{\cal G}_{\cal AA}&\mc H^{\cal G}_{\cal AB}\\
\mc H^{\cal G}_{\cal BA}&\mc H^{\cal G}_{\cal BB}\end{array}\right) T\left(\begin{array}{c} f\\f^\dagger\end{array}\right),
\ee
where the $N\times N$ matrices $\mc H_{\lambda\lambda'}^{\cal G}$ for given $\cal G$ can be read off from
Eq.~\eqref{HMajorana}, see Ref.~\cite{Pereira2020} for explicit expressions.

We next apply a unitary Bogoliubov transformation, 
\be \label{unitaryG}
\left(\begin{array}{c} f\\f^\dagger\end{array}\right)=U_{\cal G}
\left(\begin{array}{c} a\\a^\dagger\end{array}\right),
\ee
in order to diagonalize Eq.~\eqref{BdG} in terms of new (complex)
matter fermions $a_{\mu}$,
\be\label{HKa}
H_K=\frac12 \sum_{\mu=1}^{N} \varepsilon_{\mu} \left( 2a^\dagger_{\mu}a^{\phantom\dagger}_{\mu}-1\right), 
\ee
where $\varepsilon_{\mu}$ are the non-negative eigenenergies ordered as
\begin{equation}\label{ordering}
    0\le\varepsilon_{1}\le \varepsilon_{2 }\le \cdots\le \varepsilon_{N}.
\end{equation}
We often use the additional index ${\cal G}$, i.e., $a_\mu\to a_{{\cal G},\mu}$ and $\varepsilon_\mu\to \varepsilon_{{\cal G},\mu}$, to emphasize that those operators and energies 
refer to the corresponding gauge configuration. 
The matter ground state, $|\mc \varphi_0(\mc G)\rangle$, is determined by the conditions
$a_\mu|\mc \varphi_0(\mc G)\rangle=0$ (for all $\mu$) and has the energy
\begin{equation}\label{ener}
E_{{\cal G},0} = -\frac12\sum_{\mu=1}^{N}\varepsilon_{{\cal G},\mu}.
\end{equation}
However, we still have to check that this state respects the parity constraint \eqref{parityconstraint}. 
To that end, we first note that the parity of the $a$ fermions, $(-1)^{N_a}$ with $N_a=\sum_\mu a_\mu^\dagger a_\mu^{}$, 
satisfies the relation 
\be \label{parityconstraint3}
(-1)^{N_f}= (-1)^{N_a} \, {\rm det}\, U_{\cal G},
\ee 
where we have verified that the proof for Eq.~\eqref{parityconstraint3} given in Ref.~\cite{Zschocke2015} for
$\kappa=0$ can be extended to $\kappa\ne 0$. 
Equation \eqref{parityconstraint} can therefore be written as
\begin{equation}\label{parityconstraint2}
     (-1)^{N_a}\, \pi_{\cal G} = 1, \quad \pi_{\cal G} =  (-1)^{N_\chi} \, {\rm det} \, U_{\cal G} ,
\end{equation}
where the ground-state parity operator, $\pi_{\cal G}=\pm 1$, is gauge invariant.
For configurations with $\pi_{\cal G}=-1$, the matter ground state $|\varphi_0({\cal G})\rangle$ is
not in the physical subspace.  One then has to add a single fermion to the
$\varepsilon_1$ level for satisfying the parity constraint \eqref{parityconstraint2}.
The corresponding changes, 
\be\label{parityviol}
|\varphi_0({\cal G})\rangle\to a_{\mu=1}^\dagger|\varphi_0({\cal G})\rangle,\quad
E_{{\cal G}, 0}\to E_{{\cal G}, 0}+\varepsilon_1,
\ee
are implicitly understood below.  

\begin{figure*}[t]
\centering
\includegraphics[scale=0.5]{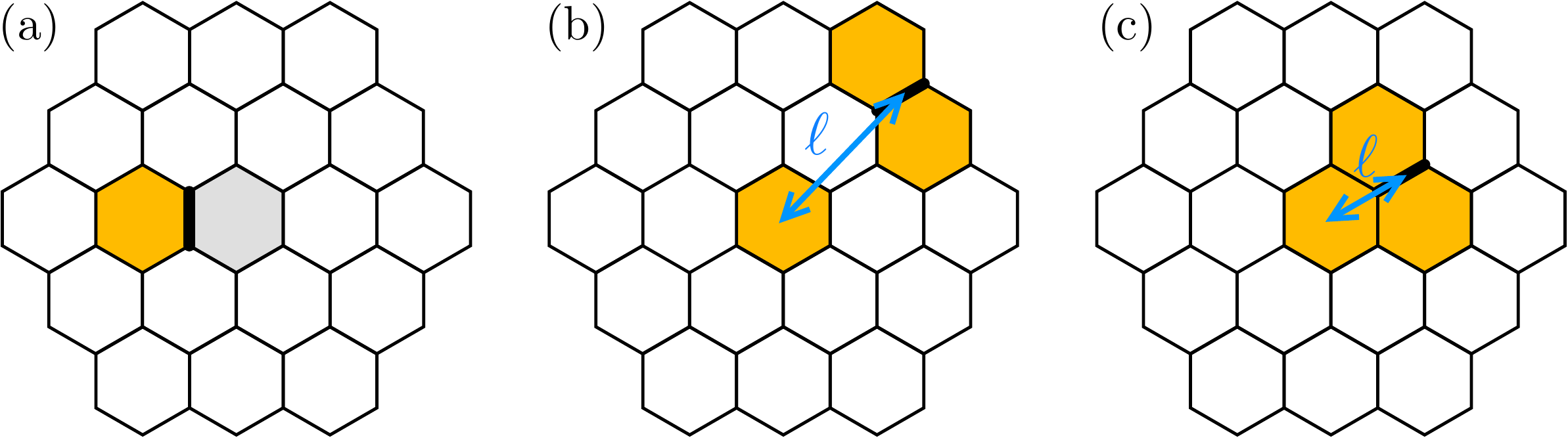}
\caption{Illustration of several gauge configurations ${\cal G}_j^{\alpha}$ contributing to 
the dynamical spin correlation functions determining the tunnel conductance.  The central plaquette
always refers to one of the two well-separated $\mathbbm{Z}_2$ vortices (the other one is not shown) 
present in the reference configuration ${\cal G}$.  
Thick black bonds indicate a flip of the corresponding bond variable 
$u_{\langle jl\rangle_\alpha}\to -u_{\langle jl\rangle_\alpha}$. 
(a) The vortex is translated by one plaquette. (b) An additional pair of adjacent vortices 
at relatively large distance $\ell$ (blue double-arrow) is created by the bond flip.   
(c)  Same as (b) but for small distance $\ell$.   
 } 
\label{fig2}
\end{figure*}

We now turn to the dynamical spin correlator, where a Fourier transformation gives the Lehmann representation (with $j$ in sublattice $\mathcal{A}$)
\be\label{Lehmann}
S_{jl}^{\alpha\beta}(\omega) =\sum_{\Phi'} \langle \Phi| \sigma_j^\alpha|\Phi'\rangle\langle\Phi'|\sigma_l^\beta|\Phi\rangle \,\delta(\omega+E_\Phi  
-E_{\Phi'}).
\ee
We consider $|\Phi\rangle$ as the matter ground state $|\varphi_0({\cal G})\rangle$
for a given gauge configuration ${\cal G}$ (which we will later choose to contain two vortices), with energy $E_0=E_{{\cal G},0}$ in Eq.~\eqref{ener}.
Inserting the Majorana decomposition into Eq.~\eqref{Lehmann}, we next observe that $c_j^\alpha$  
commutes with all terms in $H_K$ that do not contain $u_{\langle jl\rangle_\alpha}$, but
anticommutes with all terms that do. 
Starting from $\mathcal{G}=\{u_{\braket{j'l'}_{\alpha'}}\}$, we then define a new gauge 
configuration $\mathcal{G}^\alpha_j=\{\tilde{u}_{\braket{j'l'}_{\alpha'}}\}$, see Fig.~\ref{fig2}, with the bond variables 
\be
\tilde{u}_{\braket{j'l'}_{\alpha'}}=\begin{cases}-{u}_{\braket{j'l'}_{\alpha'}},& \mathrm{if}\, \,\braket{j'l'}_{\alpha'}=\braket{jl}_\alpha,\\
{u}_{\braket{j'l'}_{\alpha'}},
&\mathrm{otherwise}.\end{cases}
\ee
With this definition, Eq.\eqref{Lehmann} yields \cite{Pedrocchi2011,Knolle2014,Zschocke2015}
\begin{widetext}
\be\label{str2}
S^{\alpha\beta}_{jl}(\omega)=\sum_{\varphi(\mathcal{G}^\alpha_j)}
 \braket{\varphi_0(\mathcal{G})|c_j|\varphi(\mathcal{G}^\alpha_j)}\braket{\varphi(\mathcal{G}^\alpha_j)|c_l|\varphi_0(\mathcal{G})}\delta\left(\omega+E_0-E_{\varphi(\mathcal{G}^\alpha_j)}\right)
 \left(\delta_{jl}-iu_{\braket{jl}_\alpha}\delta_{\braket{jl}_\alpha}\right)\delta_{\alpha\beta}.
 \ee
\end{widetext}
Here $\delta_{\langle jl\rangle_\alpha}=1$ if $(jl)$ form a nearest-neighbor bond
of type $\langle jl\rangle_\alpha$, and zero otherwise.
Hence $S_{jl}^{\alpha\beta}(\omega)\ne 0$ is possible only for
equal spin indices ($\alpha=\beta$) and on-site terms or nearest-neighbor bonds.
As sketched in Fig.~\ref{fig2},  ${\cal G}$ and 
${\cal G}_j^\alpha$ are connected by either moving a vortex by one plaquette, or  by creating two
additional vortices.  We note that the zero-frequency peak in $S_G(\omega)$ 
is connected to the configurations in Fig.~\ref{fig2}(a).  Since we expect this peak 
to move to a finite but very small frequency $\omega_0$ in practice, see Sec.~\ref{sec4}, we have taken it into
account with the full weight of the $\delta$-peak in the tunnel conductance \eqref{tunnelcond}, even though
the integral in Eq.~\eqref{tunnelcond} runs over positive frequencies only.

Since  matter states for two different gauge configurations are needed in Eq.~\eqref{str2},
it is convenient to use the notations 
\bea \nonumber
 a_\mu&=&a_{\mathcal{G},\mu},\quad b_\mu=a_{\mathcal{G}^\alpha_j, \mu},\\  \label{notation}
 |0_a\rangle &=&  |\varphi_0({\cal G})\rangle,\quad |0_b\rangle= |\varphi_0({\cal G}_j^\alpha)\rangle,
\eea
with the $N$-component spinors $a=(a_1,\ldots, a_N)^T$ and $b=(b_1,\ldots, b_N)^T$.  
The $a$ matter fermions with ground state $|0_a\rangle$ thus refer to the gauge configuration ${\cal G}$, 
while the $b$ fermions with ground state $|0_b\rangle$ refer  
to ${\cal G}_j^\alpha$.  
The corresponding ground-state energies are denoted by 
$E_{|0_a\rangle}$ and $E_{|0_b\rangle}$, respectively.
From Eq.~\eqref{unitaryG},  the $a$ and $b$ fermions must be 
connected by a unitary Bogoliubov transformation \cite{Knolle2014,Zschocke2015,Blaizot1986},
\begin{equation}\label{Xdef}
\begin{pmatrix} b\\ b^\dagger \end{pmatrix}
=
{\cal W} \begin{pmatrix} a\\ a^\dagger \end{pmatrix},\quad 
{\cal W} = U_{{\cal G}_j^\alpha}^\dagger U_{\cal G}^{}=\begin{pmatrix}X^*&Y^*\\Y&X\end{pmatrix},
\end{equation}
where the $N\times N$ matrices $X$ and $Y$ satisfy the relations 
\begin{eqnarray*}
&& XX^\dagger+YY^\dagger=1,\quad X^\dagger X+Y^T Y^*=1,\\
\nonumber 
&& XY^T+YX^T=0,\quad X^TY^*+Y^\dagger X=0.
\end{eqnarray*}
For ${\rm det}\, {\cal W}=+1$, we next observe that
$|0_b\rangle$ can be obtained from $|0_a\rangle$ by means of the Thouless theorem \cite{Bertsch2009}.
As a result, one finds \cite{Zschocke2015,footnote}
\be\label{Thouless}
\ket{0_b}= [{\rm det}(X^\dagger X)]^{1/4}\, \exp\left(-\frac12 a^\dagger\, X^{\ast -1} Y^\ast\, a^\dagger\right)\ket{0_a}.
\ee
The matrix elements needed in Eq.~\eqref{str2} are of the form
\begin{equation}
\label{eq:dsf_matrixelements}
\braket{\varphi_0(\mathcal{G})|c_j|\varphi(\mathcal{G}^\alpha_j)}=\braket{0_a| c_j b^\dagger_{\mu_1}...b^\dagger_{\mu_n}|0_b}
\end{equation}
where $\mu_1\le \cdots \le \mu_n$ and $n$ is constrained by $(-1)^n=\pi_{{\cal G}_j^\alpha}$.  One can
understand this constraint by noting that Eq.~\eqref{eq:dsf_matrixelements}, which is a matrix element of the single fermion operator $c_j$, must vanish if
$|\varphi_0({\cal G})\rangle$ and $|\varphi({\cal G}_j^\alpha)\rangle$ have the same fermion parity.
We note that for ${\rm det}\,{\cal  W}=1$,  exactly one of the two 
fermionic vacua $|0_a\rangle$ and $|0_b\rangle$ will not be in the physical subspace since the $\pi_{\cal G}$ operator will change sign when flipping a bond.  
As discussed above, we therefore have to add a single fermion to one of the 
two states. Using Eq.~\eqref{Thouless} and the relation $c=TU_{\cal G} \,(a,a^\dagger)^T$,
which follows from Eqs.~\eqref{ctrans} and \eqref{unitaryG}, we can  finally 
express all matrix elements \eqref{eq:dsf_matrixelements} exclusively in terms of $a$ and $a^\dagger$ operators,
facilitating their practical computation.

For a numerical implementation, we restrict the number $n$ of excitations 
in Eq.~\eqref{eq:dsf_matrixelements} by imposing $0\le n\le n_{\rm max}$.
Under this truncation, exactness of the computed dynamical spin correlations is ensured only for 
frequencies 
\be
\omega< \omega_{\rm max}= E_{|0_b\rangle}-E_{|0_a\rangle}+\sum_{\mu=1}^{n_{\rm max}+2} 
\varepsilon_{{\cal G}_j^\alpha,\mu}.
\ee
However, already for $n_{\mathrm{max}}=2$, accurate results can be obtained even for $\omega>\omega_{\mathrm{max}}$ in the vortex-free configuration ${\cal G}_0$ \cite{Zschocke2015}.
For the two-vortex configuration ${\cal G}$, rapid convergence of the numerical results upon 
increasing $n_{\rm max}$ was observed. Since the characteristic MZM features stem from the low-frequency part of
$S_{jl}^{\alpha\beta}(\omega)$, in all cases shown here, a truncation with $n_{\rm max}=2$ was sufficient to reach 
convergence for $\omega<\omega_{\rm max}$. 

However, for selected bonds $\langle jl\rangle_\alpha$ in the two-vortex
configuration $\cal G$, we find that $\det\, {\cal W}=-1$. In such 
cases, the Thouless theorem breaks down and $X$ in Eq.~\eqref{Xdef} is a singular $N\times N$ matrix.  As a result, Eq.~\eqref{Thouless} does not apply anymore.
For computing the STS tunnel conductance near a single $\mathbb{Z}_2$ vortex, it
is essential to resolve this issue.  
For closely related problems, Refs.~\cite{Cozzini2007,Bertsch2009} have 
obtained a solution by interchanging the ground-state occupancies of a single 
particle  and its hole partner. We follow their approach and define the matrices 
$X^{(\mu)}$ and $Y^{(\mu)}$, see~Eq.~\eqref{Xdef}, according to
\begin{equation}\label{inter}
X^{(\mu)}_{kl}=\begin{cases}
X_{kl},\quad &l\neq\mu\\
Y_{kl},\quad &l=\mu
\end{cases},\quad Y^{(\mu)}_{kl}=\begin{cases}
Y_{kl},\quad &l\neq\mu\\
X_{kl},\quad &l=\mu
\end{cases},
\end{equation}
where $\mu$ refers to the index of the interchanged particle and hole.
This interchange of columns renders $X^{(\mu)}$ non-singular as it corresponds to a Bogoliubov transformation with 
positive determinant.  We can then use the Thouless theorem again, such that after the operation \eqref{inter}, 
we can effectively use Eq.~\eqref{Thouless}. The thereby obtained state, $|0'_b\rangle$, has the energy
$E_{|0'_b\rangle}=E_{{\cal G}_j^\alpha,0}+\varepsilon_{{\cal G}_j^\alpha,\mu}$, and 
the chosen index $\mu$ should  minimize $\varepsilon_{{\cal G}_j^\alpha,\mu}$.  
For instance, if it corresponds to a zero mode, $\varepsilon_{{\cal G}_j^\alpha,\mu}=0$, 
the interchange \eqref{inter} introduces no approximation, the energy ordering
in Eq.~\eqref{ordering} remains unaffected, and $|0'_b\rangle$ captures the ground state
for the $b$ fermions.  For the configurations studied in this work, we can always find a low-energy fermion 
level that approaches a zero mode in the thermodynamic limit for $\kappa\ne 0$.  These low-energy modes
are well separated from the fermion continuum which has a finite gap $\propto |\kappa|$. 

It is worth mentioning that two consistency checks are passed successfully by our numerical calculations.
First, $\lim_{V\to \infty} \int_0^{eV} d\omega\, S_{jl}^{\alpha\beta}(\omega)$ recovers the
static equal-time spin correlator \cite{Pereira2020}. 
Second, dynamical spin correlations are radially isotropic around an isolated $\mathbb{Z}_2$ vortex despite of the presence of a gauge string.   

\begin{figure*}
\centering
\includegraphics[scale=.76]{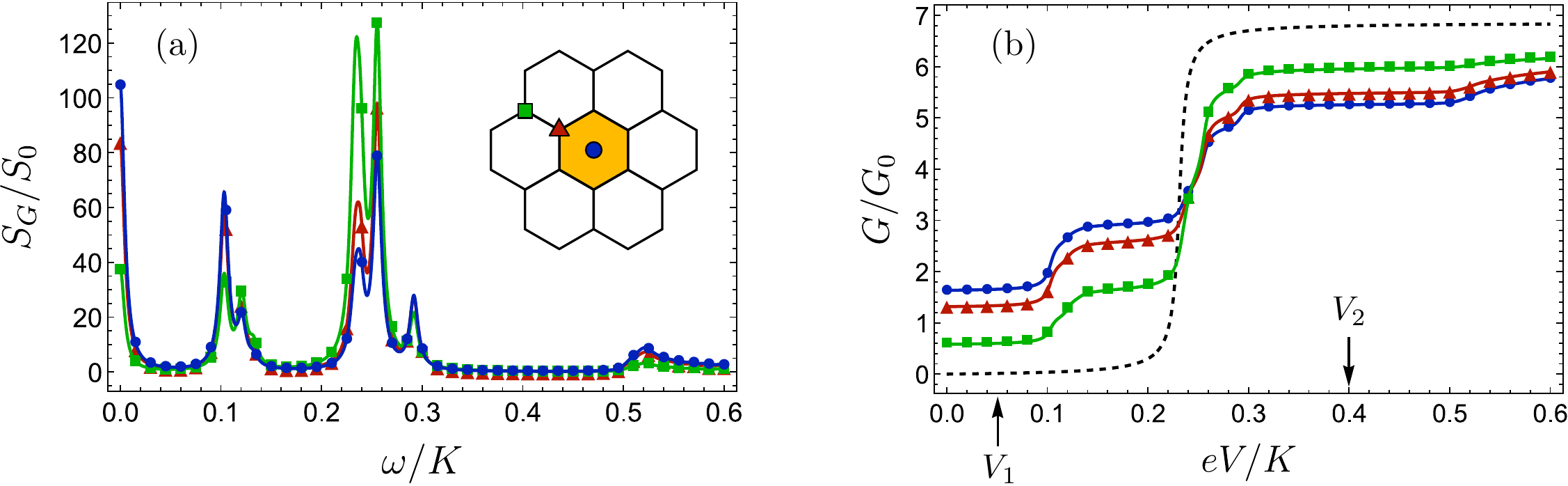}
\caption{STS for a Kitaev QSL in a two-vortex configuration $\cal G$, see Fig.~\ref{fig1}(b),
for $\kappa=0.2K$, $L=37$, $J_H=0.05 U$, $l_0=0.75a_0$, and $d=l_0$.  For $\alpha$-RuCl$_3$, one expects
$K\approx 5$~meV \cite{Winter2017b}. (a) Weighted spin correlation function $S_G$ vs $\omega$, see Eq.~\eqref{tunnelcond}, for three tip positions (inset).
We plot $S_G(\omega)$ in units of $S_0=d_Ad_B(t_0 t_B/U)^2$, with $\delta$-peaks replaced by Lorentzians of width $\Gamma_L=0.005K$ due to higher-order tunneling processes.
(b) Conductance $G$ (in units of $G_0=  S_0\frac{e^2}{\hbar}$) vs $V$, see Eq.~\eqref{tunnelcond},
for the tip positions in (a). The black dashed curve is for the vortex-free configuration ${\cal G}_0$.
The voltages $V_{1,2}$ are used in Fig.~\ref{fig4}. }
\label{fig3}
\end{figure*}

\begin{figure}
\centering
\includegraphics[scale=0.19]{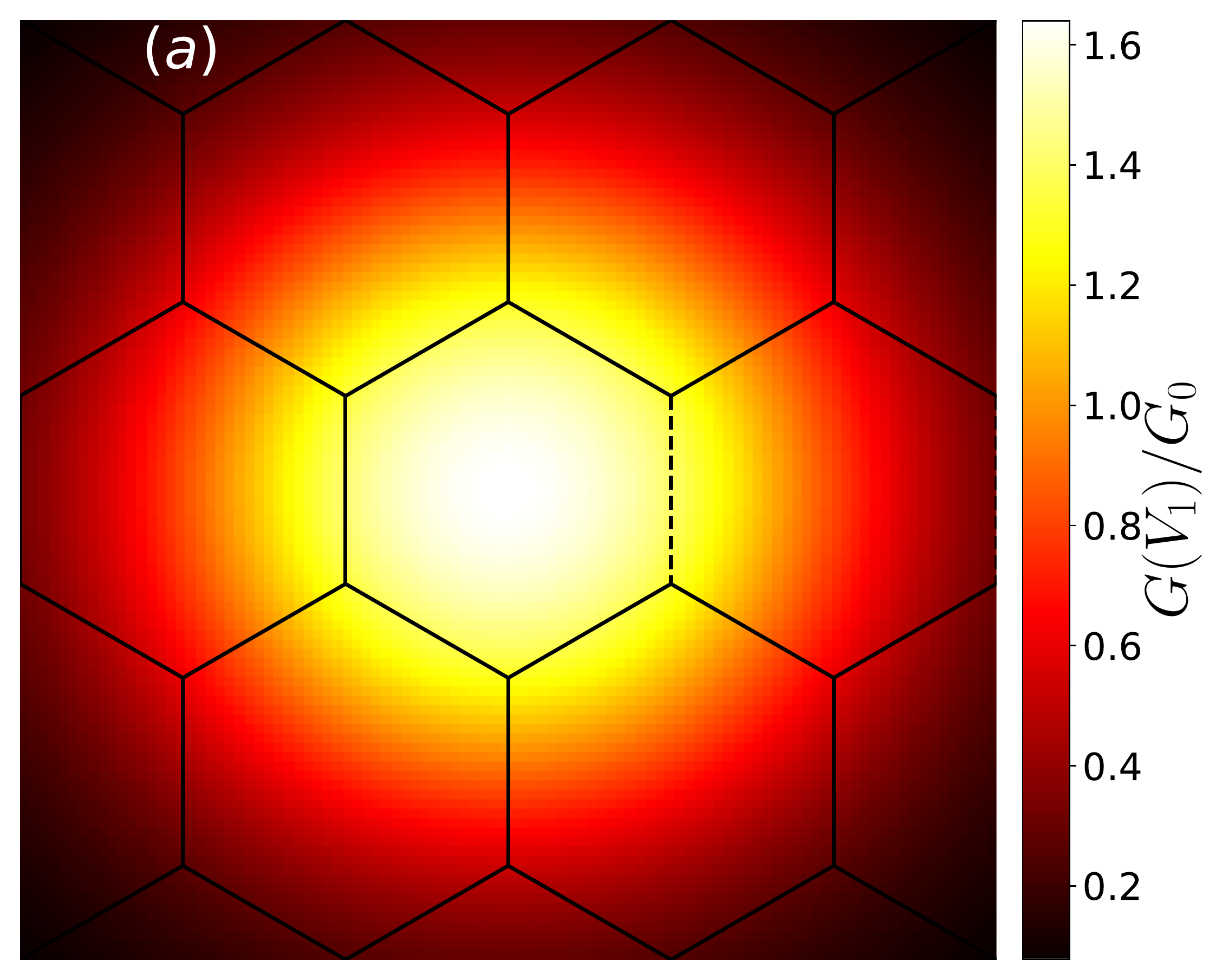}~~\includegraphics[scale=0.19]{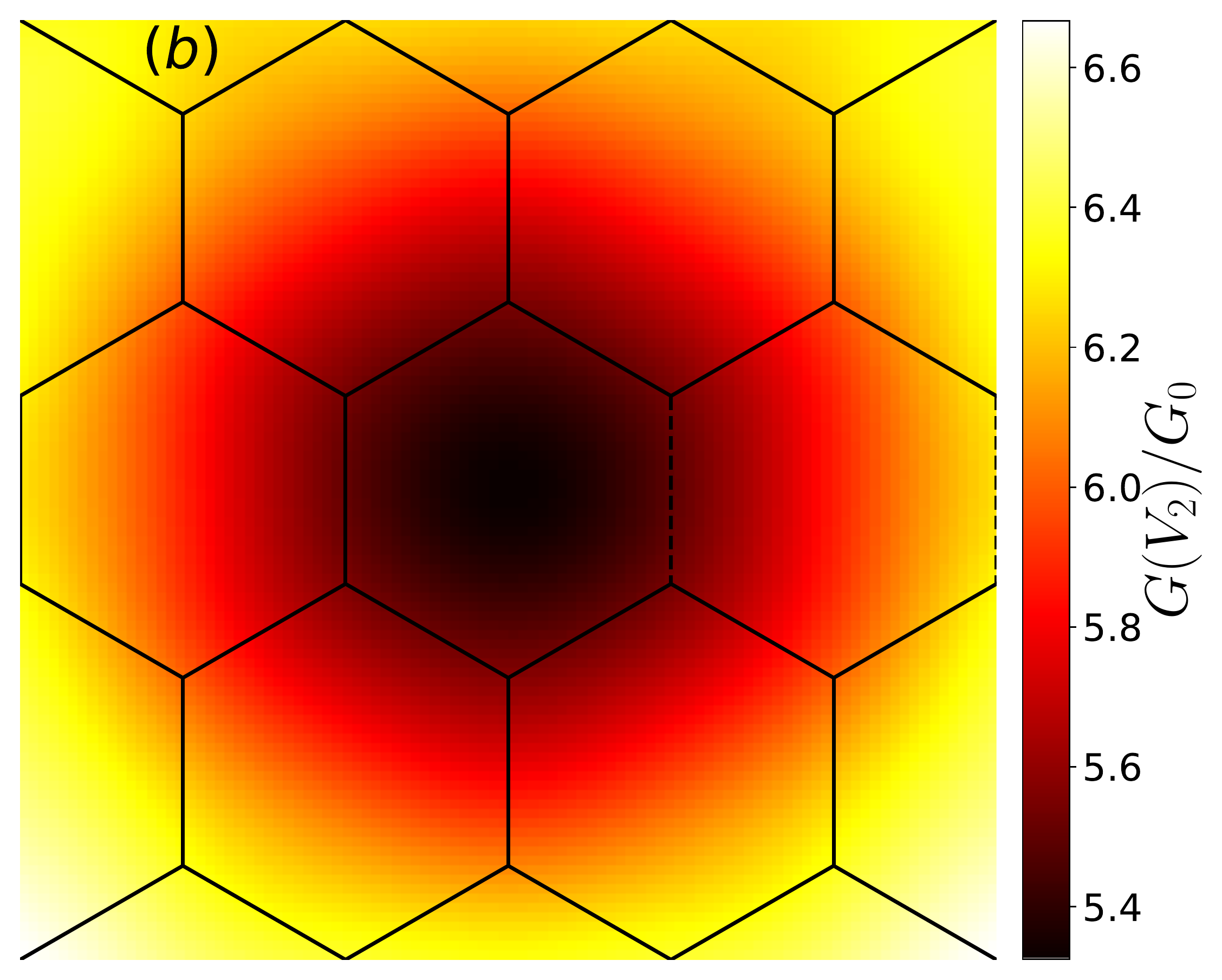}
\caption{Spatial conductance profile near a vortex (central plaquette) in the $xy$-plane, for the
parameters in Fig.~\ref{fig3} with   (a) $V=V_1$ and  (b)  $V=V_2$, see Fig.~\ref{fig3}(b).
Note the different color scales. }
\label{fig4}
\end{figure}

\section{Conductance signatures of MZMs}\label{sec4}

Figure \ref{fig3} shows numerical results for $S_G(\omega)$ and $G(V)$ for three different tip positions near an isolated $\mathbb{Z}_2$ vortex. 
The different peaks in each $S_G(\omega)$ curve have a clear physical meaning.  First, the 
$\omega=0$ peak is directly connected to MZMs and stems from configurations 
${\cal G}_j^\alpha$ with the vortex translated by one step. 
(For nonuniform Kitaev couplings, the peak can shift to a small frequency $\omega_0$, see below.)  
The support for this peak comes only from on-site terms and nearest-neighbor bonds 
directly enclosing the vortex. Indeed, Fig.~\ref{fig3}(a) shows that the peak weight decreases rapidly  with
the tip-vortex distance. Second, the peaks at $\omega=\Delta E_{2v}(\ell)~(\approx 0.1K$ in Fig.~\ref{fig3})
correspond to the energy cost for creating a configuration ${\cal G}_j^\alpha$ with an additional
pair of adjacent vortices by flipping a bond at distance $\ell$ from the original vortex, with the fermion bound state 
built from the new overlapping MZM pair unoccupied. This peak may contain several 
subpeaks since various configurations ${\cal G}_j^\alpha$ with different $\ell$, and hence different 
$\Delta E_{2v}(\ell)$, may contribute to $S_G(\omega)$ in this frequency range.  
Third, the peak structure at $\omega=\Delta E_{2v}(\ell)+\varepsilon_f(\ell)\approx 0.25K$
includes the energy cost $\varepsilon_f(\ell)$ for occupying the fermion bound state.
Finally, the onset of the gapped two-fermion continuum is marked by a (small) peak at 
$\omega=\Delta E_{2f}= \frac{3\sqrt3}{2}|\kappa|~(\approx 0.5K$ in Fig.~\ref{fig3}).

The conductance $G(V)$ in Fig.~\ref{fig3}(b) follows by integrating $S_G(\omega)$ 
and  therefore shows steps at the voltages matching a peak in $S_G(\omega)$.  One can thus
measure the important energy scales $\Delta E_{2v}$, $\varepsilon_f$, and $\Delta E_{2f}$ by STS.
However, the respective step sizes are not universal because the peak weights in $S_G(\omega)$ 
depend on the tip position and on the form factors.  It is instructive to compare to the
vortex-free configuration ${\cal G}_0$, see Fig.~\ref{fig3}(b), where $G(V)$ is strongly suppressed  
for $eV<\Delta E_{2v}(\infty)+\varepsilon_f(\infty)$. Indeed, here the lowest-energy excitation probed by $G(V)$  
corresponds to adding a vortex pair and filling the fermion bound state in order to respect the parity constraint.  
In this low-voltage regime, the conductance for the two-vortex configuration ${\cal G}$ is instead dominated by 
 MZMs and will be finite at small $V$, with a step at $eV=\Delta E_{2v}(\ell)$. 
We also observe from Fig.~\ref{fig3}(b)  that the ``bulk'' behavior of $G(V)$, found for arbitrary
tip position in configuration ${\cal G}_0$, is approached by moving the probe tip far away from the vortex center.
We note that the zero-voltage step is particular to the integrable Kitaev model with 
uniform couplings (assumed in Fig.~\ref{fig3}), where the eigenstates are degenerate with respect 
to the vortex position. In a generic nonintegrable case, vortices are mobile but can be trapped
by bond disorder, vacancies, magnetic impurities, or by an external electrostatic potential. 
The $V=0$ step may then shift to a small finite voltage $eV=\omega_0$, 
where $\omega_0$ describes the difference in vortex creation energies on different plaquettes.
Such shifts may be useful to distinguish MZM-induced conductance steps 
from the background conductance due to $T_0$ in Eq.~\eqref{Hcot}.

For the voltages $V_{1,2}$ marked in Fig.~\ref{fig3}(b),  we show the tip-position dependence of 
the conductance in Fig.~\ref{fig4}. 
For $V=V_1$, see Fig.~\ref{fig4}(a), the physics is dominated by the 
zero-frequency MZM peak in 
$S_G(\omega)$, and the spatial profile in Fig.~\ref{fig4}(a) encodes a convolution
of the (squared) MZM wave function \cite{Hassler2019} with the form factor \eqref{cab}. 
However, in contrast to the standard situation in STS \cite{STMreview}, 
it is not possible to map out the MZM wave function beyond the immediate vicinity of the vortex
because only terms from sites or bonds encircling the vortex  contribute for $eV<\Delta E_{2v}(\ell)$.
The conductance profile for $V=V_2$ in Fig.~\ref{fig4}(b) reveals a dip in the center,
which arises because for a tip away from the vortex, 
the form factors enhance the peak contribution for $eV>\Delta E_{2v}(\ell)$.
However, this voltage regime involves many vortex configurations ${\cal G}_j^\alpha$,
rendering it difficult to extract the MZM wave function.
Nonetheless, the conductance profile allows to detect the MZM at the vortex location. 
Finally, the angular isotropy  of the spatial profile approximately found at low voltage
is reduced to a $C_6$ symmetry at higher voltages.  
While this effect is hardly visible for the tip distance $d=l_0$
in Fig.~\ref{fig4}, it becomes more prominent for smaller $d$.

\section{Topological magnons}\label{sec5}

In this section, we explore a different mechanism that could in principle 
generate similar tunnel conductance features as those reported above for MZMs in the spin-liquid phase.
To that end, we consider topological magnons in the polarized phase of the Kitaev model in 
a magnetic field \cite{Kim1,Kim2,Feldmeier2020}.  Such models
have been proposed   as alternative scenario for explaining the 
observed half-quantized thermal Hall conductivity \cite{Kim1,Kim2}.
Below we clarify whether local defects or magnetic impurities are able to generate 
 topological magnon bound states below the magnon gap.  If present, such bound states may produce tunnel conductance steps at voltages matching the respective bound-state energies. 
In analogy to the topological superconductor case, 
magnon-induced conductance steps could then be difficult to distinguish from those
caused by MZMs in a Kitaev spin liquid. 

We consider spin-$S$ operators $S_i^\gamma$ on the 2D honeycomb lattice with Kitaev couplings.
The Hamiltonian is given by
\be\label{Hmagnon}
H_m =-\sum_{\langle ij\rangle_\gamma } K_{ij} S^\gamma_i S^\gamma_j - \sum_j\mb h_j\cdot  \mb S_j,
\ee
where $\gamma\in \{x,y,z\}\equiv \{1,2,3\}$ denotes the spin components as well as the bond directions,
see Sec.~\ref{sec2}.
For simplicity, we assume that the local magnetic fields are oriented along the $[111]$ direction,
$\mb h_j=h_j \mb c$, with the unit vector $\mb c$ in Eq.~\eqref{axes}, see App.~\ref{appA}. 
In the homogeneous case, the Kitaev couplings and local fields are given by $K_{ij}=K$ and $\mb h_j=\mb h$, 
respectively.
In order to model a \emph{defect}, we study inhomogeneous Kitaev couplings $K_{ij}$ near a single plaquette corresponding to the defect, similar to models for bond disorder and vacancies \cite{Knolle2019b,Kao2021,Dantas2021}.  
Recalling that a large-spin \emph{magnetic impurity} is equivalent to a local change of the 
magnetic field at a single site \cite{Imry1975}, we model a  magnetic impurity  by a 
local change of the field $h_i\ne h$ at this site relative to the bulk field $h$. 
We follow Refs.~\cite{Kim1,Kim2} and derive the linear spin wave theory  which 
becomes exact in the large-$S$ limit. 

We first rotate the local basis to have the magnetization axis along the $c$ direction. With the orthogonal matrix
$R=(\mb a\, \mb b\, \mb c)$, see Eq.~\eqref{axes}, we have the rotated spin operators 
$\tilde S^\alpha_i=R_{\alpha\beta } S^\beta_i$. Next, we employ a Holstein-Primakoff transformation to expand around the
 polarized state, 
\bea\nonumber
\tilde {S}_i^z&=&S-b^\dagger_i b^{\phantom\dagger}_i,\quad
\tilde {S}_i^x\approx \sqrt{\frac{S}{2}}(b^{\phantom\dagger}_i+b^\dagger_i),\\
\tilde {S}_i^y&\approx& -i\sqrt{\frac{S}{2}}(b^{\phantom\dagger}_i-b^\dagger_i),
\eea
with bosonic magnon operators $b_i$.  Expanding $H_m$ in Eq.~\eqref{Hmagnon} in powers of $1/S$, we obtain 
$H_m = E_{\rm cl}+H_1+H_2+\mc O(S^{1/2}).$
The first term describes the classical ground state energy,  $E_{\rm cl}=- \frac{S^2}3\sum_{\langle ij\rangle}K_{ij}-S\sum_j h_j$. The second term is linear in the bosons,
\be\label{H1}
H_1=\frac{S^{3/2}}{3}\sum_i \left(\sum_\gamma e^{-i2\pi \gamma/3}K_{i,i+\boldsymbol \delta_\gamma}\right)b_i^{\phantom\dagger} +\text{h.c.},
\ee
with the in-plane nearest-neighbor vectors 
\be
\boldsymbol\delta_1=\frac12 \hat{\mb x}+\frac{1}{2\sqrt3}\hat{\mb y},\quad
\boldsymbol\delta_2=-\frac12 \hat{\mb x}+\frac{1}{2\sqrt3}\hat{\mb y},\quad
\boldsymbol\delta_3=-\frac{1}{\sqrt3}\hat{\mb y}.
\ee
One finds $H_1=0$ for $K_{ij}=K$, but in the presence of defects, $H_1\ne 0$  
indicates that we have expanded around the wrong classical state. Due to the anisotropy
of the Kitaev interactions, the spins do not align with the [111] direction anymore if the $\mathbb Z_3$ symmetry is
broken by defect bonds. To correct for this problem, one has to find the correct classical state with an 
inhomogeneous magnetization and then apply position-dependent $R$ matrices in order to rotate the spins 
to their local magnetization axis.  While such refinements could give quantitative corrections, we here 
focus on the quadratic term,
\begin{widetext}
\be
H_2=-\frac{S}3\sum_{\langle ij \rangle_\gamma}K_{ij}\left(   b^{ \dagger}_ib^{\phantom\dagger}_j+b^{ \dagger}_jb^{\phantom\dagger}_i + e^{i2\pi \gamma/3} b^{\phantom\dagger}_ib^{\phantom\dagger}_j+e^{-i2\pi \gamma/3} b^{ \dagger}_j b^{ \dagger}_i \right)+S\sum_i  \left(h_i +\frac13\sum_j K_{ij}\right) b^{ \dagger}_ib^{\phantom\dagger}_i.\label{H2}
\ee
Indeed, in general terms, the linear spin wave theory resulting from Kitaev (or other) interactions on the 2D 
honeycomb lattice must be of the form
\be\label{gen22}
H_2=S\sum_{\langle ij \rangle_\gamma}\left( t_{ij}  b^{ \dagger}_ib^{\phantom\dagger}_j+t_{ij}^*b^{ \dagger}_jb^{\phantom\dagger}_i + \Delta_{ij}b^{\phantom\dagger}_ib^{\phantom\dagger}_j+\Delta_{ij}^*b^{ \dagger}_jb^{ \dagger}_i \right)+S\sum_i   B_i\, b^{ \dagger}_ib^{\phantom\dagger}_i,
\ee
\end{widetext}
where $B_i$ is an effective magnetic field including the Weiss field. 
The misalignment of spins around defects here should give rise to an additional position dependence 
in the parameters $t_{ij}$, $\Delta_{ij}$ and $B_i$ in Eq.~\eqref{gen22}, on top of  the 
immediate  effects of $K_{ij}$-anisotropy in Eq.~(\ref{H2}). 
In what follows, we consider $H_m\simeq H_2$ as given by Eq.~\eqref{H2}.

\begin{figure}
\begin{center}
\includegraphics[width=.95\columnwidth]{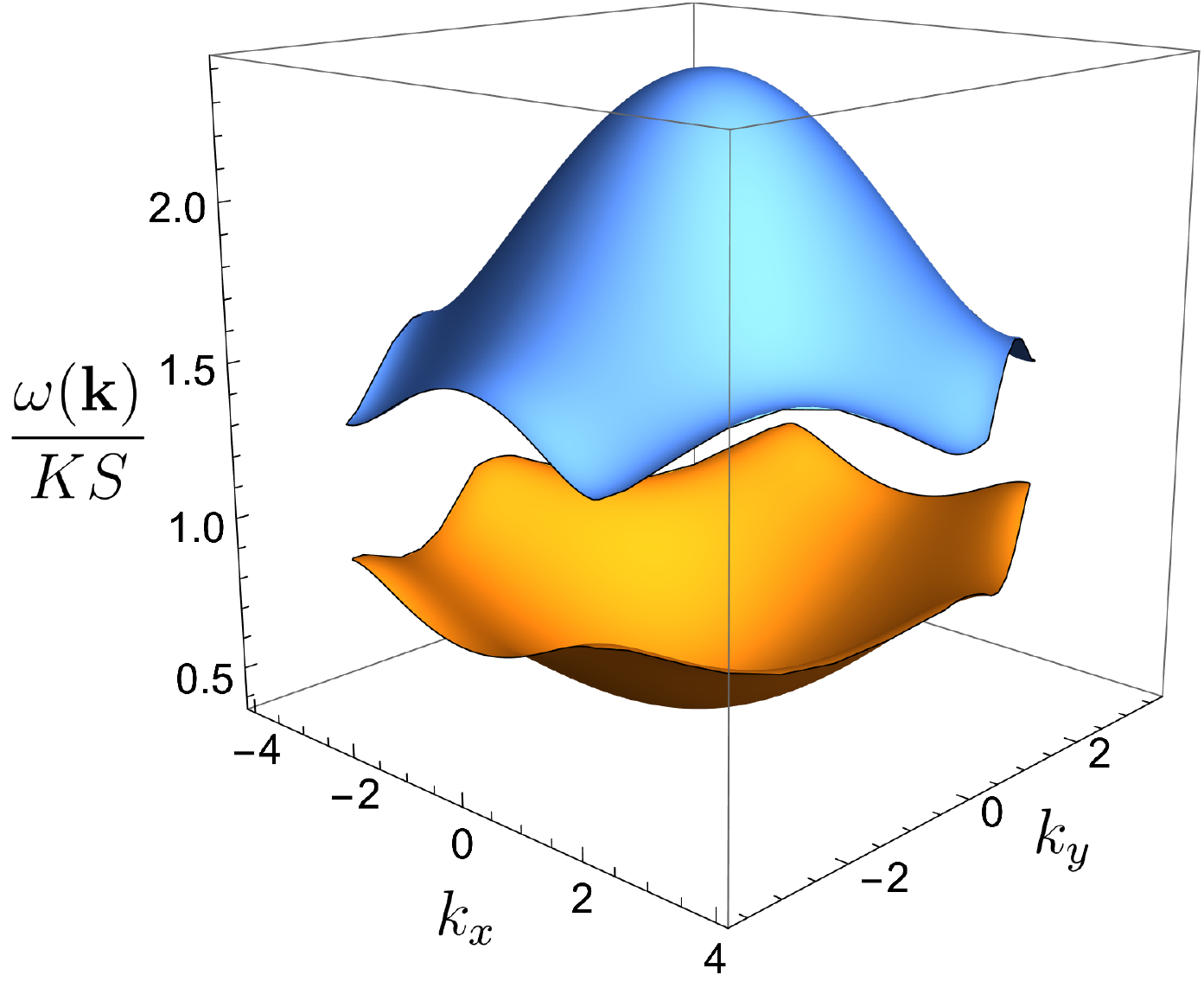}
\end{center}
\caption{Topological magnon bands for $h/K=0.4$ (with momentum unit $a_0^{-1}$) 
from linear spin wave theory for the homogeneous model~\eqref{Hmagnon}. }
\label{fig5}
\end{figure}

We first address the homogeneous case, where Fourier transformation gives
$H_2=  S \sum_{\mb k\in \frac12\text{BZ}}\Psi^\dagger_{\mb k}M_{\mb k}
\Psi^{\phantom\dagger}_{\mb k}$. Here $\mb k$ runs over half the Brillouin zone,
 $\Psi^\dagger_{\mb k}= (\begin{array}{cccc}b^\dagger_{\mb k,{\cal A}}&b^\dagger_{\mb k,{\cal B}}&b^{\phantom\dagger}_{-\mb k,{\cal A}}&b^{\phantom\dagger}_{-\mb k,{\cal B}}
\end{array})$
is a four-component spinor (including the sublattice index), and 
\be
M_{\mb k}=\left(\begin{array}{cc}A_{\mb k}&B_{\mb k}\\
B^*_{-\mb k}&A^{\rm T}_{-\mb k}
\end{array}\right).
\ee 
Using the notation $\Gamma_{\mb k, n}=\sum_{\gamma} e^{-i2\pi n\gamma /3}e^{i\mb k\cdot \boldsymbol\delta_\gamma}$ with $n\in \{0,1\}$, we have defined the matrices 
\bea
\nonumber A_{\mb k}&=&\left(\begin{array}{cc}
h+K&-\frac13K\Gamma_{\mb k,0}\\
-\frac13K\Gamma_{-\mb k,0}&h+K
\end{array}\right),\\
B_{\mb k}&=&\left(\begin{array}{cc}
0&-\frac13K\Gamma_{\mb k,1}\\
-\frac13K\Gamma_{-\mb k,1}&0
\end{array}\right).
\eea
This Hamiltonian can be diagonalized by a Bogoliubov transformation. With $\Sigma=\text{diag}(1,1,-1,-1)$, we obtain the 
magnon band dispersion from  the positive eigenvalues of $\Sigma M_{\mb k}$. The result is illustrated in Fig.~\ref{fig5}. 
We find two bands $\omega_1(\mb k)$ and $\omega_2(\mb k)$, where analytical but 
lengthy expressions are available.  These topological magnon bands cover the energy range
\be\label{bands}
h<\omega_1(\mb k) < \sqrt{h(h+2K)},\quad h+K <\omega_2(\mb k) <  h+2K.
\ee
The magnon band gap is thus given by $\Delta E_m=h$.
For $h\to 0$, the lower magnon band becomes a zero-energy flat band, signalling the 
degeneracy of the classical Kitaev model at zero field.  

\subsection{Defect from bond disorder}

Next we turn to inhomogeneous Kitaev interactions, where we model a defect by modifying the bonds $K_{ij}\to \xi K$ around a given plaquette representing the defect by a positive factor $\xi\ne 1$.  
We have studied two different radially symmetric bond defect  patterns.  In the first case, we modify only the six bonds directly
surrounding the defect plaquette.  In the second case, we instead change only the six adjacent bonds pointing radially outward from this plaquette.  
The conclusions described below are identical for both cases.
We have studied the spectrum of $H_2$ in Eq.~\eqref{H2} by numerical diagonalization on a finite $L\times L$ honeycomb lattice as described in Sec.~\ref{sec3}.  We observe that 
making the bonds stronger ($\xi>1$) creates a repulsive potential for magnons,
which generates anti-bound states above the top of the upper band, $\varepsilon_m'>h+2K$. 
There are also bound states in the gap between both bands.  However, even if we make the bonds significantly weaker, $\xi<1$, we never observe bound states below the lower band, $\varepsilon_m<\Delta E_m$.  We conclude that bond defects are unlikely to produce magnon bound states at subgap energies. 
At the same time, we cannot rule out that a more complex bond defect pattern could cause subgap features that can mimic the Majorana features described in Sec.~\ref{sec4}.  Future work should
investigate this issue in more detail.

\begin{figure}
\begin{center}
\includegraphics[width=0.9\columnwidth]{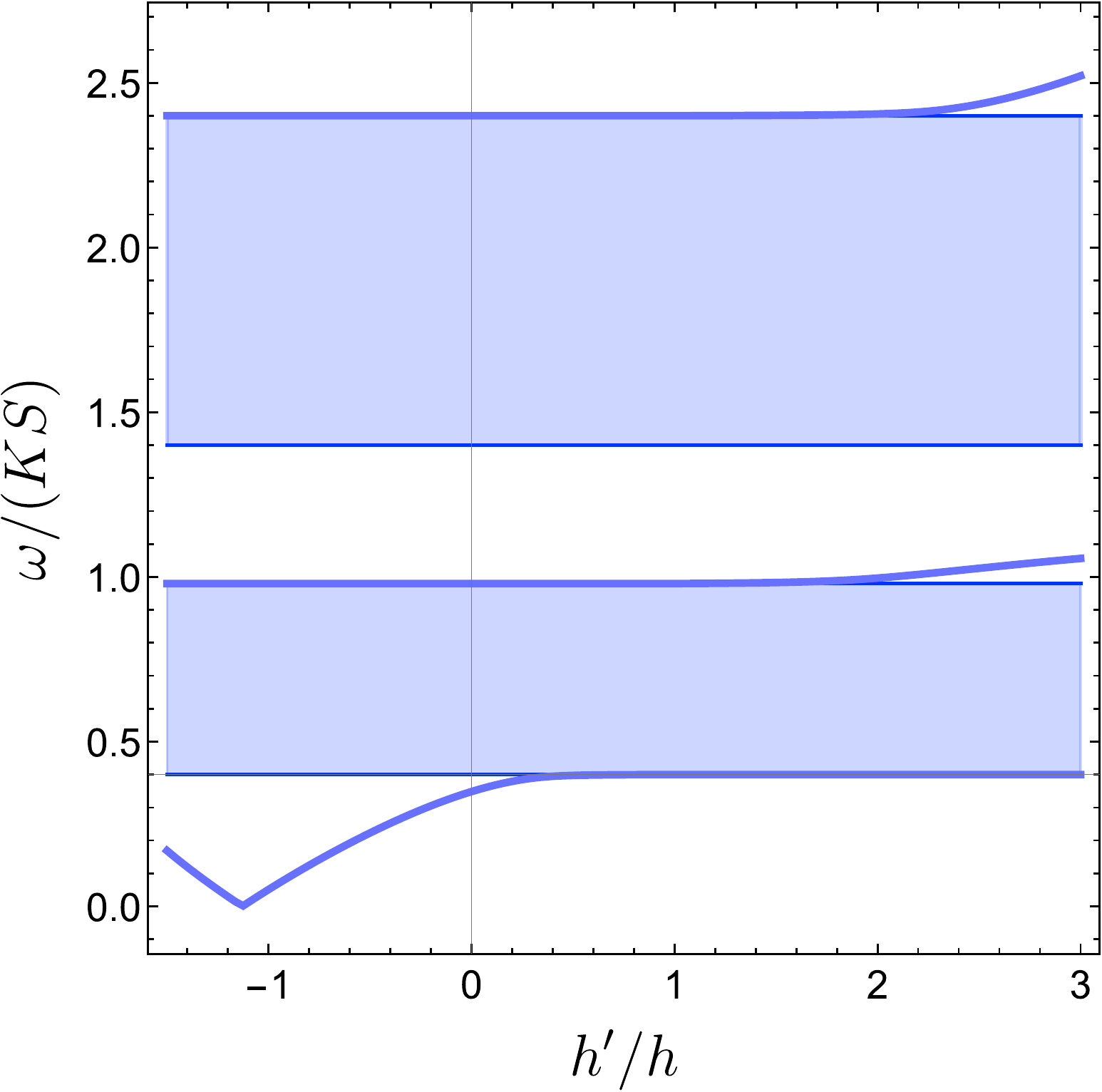}
\end{center}
\caption{Magnon spectrum $\omega$ vs $h'/h$ for a local magnetic field $h'\ne h$ at a single site. 
The  bulk field is $h=0.4K$. Shaded regions describe continuum states, see Eq.~\eqref{bands}.
A single subgap bound state can exist for $h'<h$. A high-energy anti-bound state is visible for $h'>h$, 
and another bound state exists in the minigap between both bands.
}
\label{fig6}
\end{figure}

\subsection{Magnetic impurity}

Another limiting case is to locally modify only the magnetic field $h_i$ in Eq.~\eqref{H2}, 
keeping homogeneous Kitaev couplings $K_{ij}=K$. 
For a radially symmetric inhomogeneous magnetic field profile,
$\mathbb{Z}_3$ symmetry remains intact and the linear-boson term $H_1$ in Eq.~\eqref{H1} vanishes.  
If we change the field only at a single site, $h_i= h'\ne h$, with the
bulk field $h$ acting at all other sites, we can find a single subgap bound state for $h'<h$ as shown 
in Fig.~\ref{fig6}.  The bound-state energy $\varepsilon_m<\Delta E_m$  
vanishes for $h'\approx -1.1h$ for $h=0.4K$. For smaller $h$, the vanishing of $\varepsilon_m$ occurs at lower values of $h'/h<0$.  
For generic values of $h'/h$, we find that $\varepsilon_m$ is positive.
The dynamical spin correlation function then will have a peak at $\omega=\varepsilon_m$, 
and Eq.~\eqref{tunnelcond} yields a single step-like feature in $G(V)$ at $eV=\varepsilon_m$.  Except for the
fine-tuned case with $\varepsilon_m=0$, this step does not occur at zero voltage as
expected for the MZM case. 

For a wider field profile, with the field change extending over several sites, 
we typically find several subgap bound states.  This case can be realized if the impurity is coupled to several sites.
In such cases, from the $G(V)$ curve alone, it can be difficult to disentangle
the effects of magnon bound states from those due to MZMs.  However, a
collection of several nearby magnetic impurities causing such a field profile should 
be identifiable by concomitant STM surface topography scans. 

\section{Conclusions}\label{sec6}

Based on the above analysis, we expect that the tunnel conductance features due 
to MZMs in a spin liquid will be quite robust.
For the topological magnon scenario  in Sec.~\ref{sec5}, we find that defects modeled by locally 
inhomogeneous Kitaev couplings do not bind subgap magnon bound states.
On the other hand, a large-spin magnetic impurity can induce a single 
subgap bound state centered at the corresponding site.
One then expects a single conductance step, where the spatial distribution of the STS tunnel
conductance peaks at this site.  For the MZM case, 
we instead predict a characteristic sequence of steps and the spatial distribution 
should peak at the center of the hexagon defining the vortex.

We conclude that the perspectives for STS detection of MZMs in spin liquids appear 
promising.  In fact, tunneling experiments on monolayers of $\alpha$-RuCl$_3$ have recently observed 
interesting low-energy excitations \cite{Yang2022}. 
Given the rapid progress  in encapsulating and probing atomically thin materials \cite{Rhodes2019}, 
detailed experimental tests of our predictions will likely soon be available.

\begin{acknowledgments}  
We acknowledge funding by the Deutsche Forschungsgemeinschaft (DFG, German Research Foundation),
Projektnummer 277101999 - TRR 183 (project B04), Normalverfahren Projektnummer EG 96-13/1, 
and under Germany's Excellence Strategy - Cluster of Excellence Matter 
and Light for Quantum Computing (ML4Q) EXC 2004/1 - 390534769, by the Brazilian ministries MEC and MCTI, by the Brazilian agency CNPq, and by the 
Coordena\c{c}\~{a}o de Aperfei\c{c}oamento de Pessoal de N{\'i}vel Superior - Brasil (CAPES) - Finance Code 001.
\end{acknowledgments}

\appendix

\section{Derivation of low-energy theory }\label{appA}

This appendix provides a derivation of the cotunneling Hamiltonian \eqref{Hcot} with the corresponding transition matrix 
elements.  As starting point, we take the general Hamiltonian 
$
    H_{\rm tot}=H_M+V_{\rm at}+H_{\rm tun},
$
where $H_M$ describes noninteracting metallic leads representing the scanning probe tip and the substrate,
\be\label{HM}
H_M=\sum_{\nu\in \{A,B\}}\sum_{ \tau\in\{\uparrow,\downarrow\}}\sum_{\mb  k}  \varepsilon_{\nu \tau}(\mb k)  c^\dagger_{\nu  \tau}(\mb k)c^{\phantom\dagger}_{\nu \tau}(\mb k).
\ee
The fermion annihilation operators $c_{\nu\tau}(\mb k)$ with $\nu=A,B$ refer to tip and substrate electrons, 
respectively, where $\tau$ is the spin projection and $\varepsilon_{\nu\tau}(\mb k)$ the energy with respect to the Fermi energy.  
The Pauli matrices $\boldsymbol \tau$ used below act in the spin space of the conduction electrons.

For the 2D Kitaev layer, we start from a Hubbard-Kanamori model for the $d^5$ electrons in an edge-sharing octahedral environment,
e.g., those of the Ru$^{3+}$ ions in $\alpha$-RuCl$_3$.
For lowest-order perturbation theory in the tunnel Hamiltonian $H_{\rm tun}$ 
connecting the layer to the STM tip and to the substrate, only 
the single-site atomic Hamiltonian $V_{\rm at}$ in the Hubbard-Kanamori model is needed  (see, for instance,  Ref.~\cite{Pereira2020}),
\be\label{Vat}
V_{\rm at}= \frac{U-3J_H}2(\bar N-1)^2-2J_H\bar{\mb S}^2-\frac{J_H}2\bar{\mb L}^2 +\lambda_{\rm so} \bar{\mb L}\cdot
\bar{\mb S},
\ee
with the on-site Coulomb energy $U$, the Hund coupling $J_H$, and the spin-orbit coupling $\lambda_{\rm so}$. 
The respective couplings can be renormalized by screening processes resulting from the presence of the tip and the substrate, but one expects $U\approx 2$~eV and $\lambda_{\rm so}\ll J_H, U$.  For definiteness, we assume $J_H\ll U$. To lowest order in $H_{\rm tun}$, contributions from different lattice sites
simply add up.
The operators $\bar N$, $\bar{\mb S}$ and $\bar{\mb L}$ in Eq.~\eqref{Vat} refer to hole number, spin, and angular momentum, respectively. In terms of the
hole annihilation operators $h_s$ with the combined spin-orbital index $s=(\alpha,\sigma)$, they are expressed as
\be
\bar N= h^\dagger  h^{\phantom\dagger},\quad \bar{\mb S}=\frac12   h^\dagger ( \bar{\boldsymbol \sigma} \otimes \mathbbm 1_3)h^{\phantom\dagger},\quad\bar{\mb L}=  h^\dagger  (\mathbbm 1_2\otimes \bar{\mb l})h^{\phantom\dagger}, 
\ee
with $h^\dagger=(h^\dagger_{x\uparrow},h^\dagger_{y\uparrow},h^\dagger_{z\uparrow},h^\dagger_{x\downarrow},h^\dagger_{y\downarrow},h^\dagger_{z\downarrow})$.
The five $d$-electrons in a cubic crystal field occupy three $t_{2g}$-orbitals $(xy,yz,zx)$,
denoted here by the complementary index $\alpha=(z,x,y)$.  The Pauli matrices $\bar{\boldsymbol\sigma}$ act
in the spin space of the magnetic layer site, and 
$\bar{\mb l}=(\bar{l}^x,\bar{l}^y,\bar{l}^z)$ represents the $l_{\text{eff}}=1$ orbital angular momentum of the corresponding $t_{2g}$ states, with explicit matrix representations specified in Ref.~\cite{Pereira2020}. Following standard practice, the spin-orbit coupling $\lambda_{\rm so}$ will be taken into account later through a projection to the lowest-lying  hole states with total angular momentum $j_{\rm eff}=1/2$.

Electron transfer  between tip (or substrate) and the Mott insulating site is described by a tunneling Hamiltonian $H_{\rm tun}={\cal T}_1+{\cal T}_{-1}$, where ${\cal T}_{\pm 1}$ refers to changes of the hole number by $\Delta \bar N=\pm 1$, respectively.  
With the complex-valued tunnel amplitude $t_{\nu \tau s}(\mb k)$ connecting a conduction electron in lead $\nu=A,B$ with spin $\tau$ and momentum $\mb k$ to the spin-orbital hole state $s=(\alpha,-\sigma)$ on the magnetic site,
\be
{\cal T}_1=\sum_{\nu, \tau, s}\sum_{\mb k}  t_{\nu\tau s}(\mb k)  c^\dagger_{\nu\tau}(\mb k)h^\dagger_{s},
\quad {\cal T}_{-1}^{}={\cal T}_1^\dagger.
\ee 
We then employ $H_0=H_M+V_{\rm at}$ as the unperturbed Hamiltonian.
The ground-state sector has a single hole at the spin-liquid site, and
the intermediate states have either $\bar N=0$ or $\bar N=2$ holes, 
depending on whether ${\cal T}_{-1}$ or ${\cal T}_{1}$ is applied to a single-hole state. 
In the latter case, we have to distinguish between angular momentum channels with  $L=0,1,2$. 
Following Ref.~\cite{Pereira2020}, we use the notation $\mc P_{L}^{(n)}$ for the 
projection operators to states with angular momentum $L$ and hole number $n=0,1,2$.  We omit the lower index 
for $n=0,1$ because in those cases there is only a single angular momentum channel. The projector to two-hole states is  $\mc P^{(2)}=\sum_L\mc P^{(2)}_L$. For a lowest-order expansion in $H_{\rm tun}$, 
the Hilbert space can be truncated to have at most two holes at the magnetic layer site, 
 $\mathbbm 1\simeq \mc P^{(0)}+\mc P^{(1)}+\mc P^{(2)}$. 
 
Next we employ a canonical transformation to perform the projection to the low-energy sector, which is 
equivalent to a Schrieffer-Wolff transformation.  Writing $\tilde H=e^S H e^{-S}=H+[S,H]+\cdots$, 
the first-order generator $S=S_1$ must then obey $[H_0,S_1]=H_{\rm tun}$. Using the commutators 
\begin{widetext}
\begin{eqnarray}\nonumber
  && [ H_0,\mc P^{(2)}_L{\cal T}_{1}\mc P^{(1)} ] =\sum_{\nu, \tau, s}\sum_{\mb k} t_{\nu \tau s}(\mb k)  [\Delta E_{L}+\varepsilon_{\nu \tau}(\mb k)] \, c^\dagger_{\nu  \tau}(\mb k)\mc P^{(2)}_Lh^\dagger_{ s}\mc P^{(1)}, \\
&& [ H_0,\mc P^{(0)}{\cal T}_{-1}\mc P^{(1)} ] =\sum_{\nu, \tau, s}\sum_{\mb k} t^*_{\nu \tau s}(\mb k) [\Delta E_{0}-\varepsilon_{\nu \tau}(\mb k)] \, \mc P^{(0)}h^{\phantom\dagger}_{ s}\mc 
 P^{(1)}c^{\phantom\dagger}_{\nu  \tau}(\mb k),
\end{eqnarray}
and writing $S_1=S_1^{(+)}-S_1^{(-)}$ with $S_1^{(-)}=S_1^{(+) \dagger}$, the part increasing the hole number at the magnetic site is
\be
S_1^{(+)}=\sum_{\nu, \tau, s}\sum_{\mb k} c^\dagger_{\nu  \tau}(\mb k) \left(-\frac{t_{\nu \tau s}(\mb k) }{\Delta E_{0}-\varepsilon_{\nu \tau}(\mb k)}\mc P^{(1)}h^{ \dagger}_{ s}\mc P^{(0)}+
\sum_L\frac{t_{\nu \tau s}(\mb k) } {\Delta E_{L}+\varepsilon_{\nu \tau}(\mb k)}\mc P^{(2)}_Lh^\dagger_{ s}\mc P^{(1)}\right).
\ee
\end{widetext}
The excitation energies $\Delta E_L$ are given by
\be\label{DeltaEL}
\Delta E_{0}=\frac{U}{2}+J_H,\quad \Delta E_{1}=\frac{U}{2}-4J_H,\quad \Delta E_{2}=\frac{U}{2}-2J_H,
\ee
where the energy for the transition to a state with zero holes is the same as for the transition to two holes with $L=0$. The charge gap is set by the smallest of those energies, $E_g=\Delta E_1$.
The canonical transformation then results in the cotunneling Hamiltonian
\be
H_{\rm cot}=-\frac12\mc P^{(1)} \left({\cal T}_{-1}S_1^{(+)}-{\cal T}_1S_1^{(-)}\right)\mc P^{(1)}+\text{h.c.},
\ee
which accurately describes the low-energy subspace with energy scales below $E_g$.
Inserting the above expressions, we find the explicit representation
\begin{widetext}
\bea
H_{\rm cot}&=&-\frac12 \sum_{\nu_1, \tau_1, s_1} \sum_{\nu_2, \tau_2, s_2} \sum_{\mb k_1,\mb k_2} \frac{ t_{\nu_2 \tau_2 s_2}(\mb k_2)t^*_{\nu_1 \tau_1 s_1}(\mb k_1)}{\Delta E_{0}-\varepsilon_{\nu_1 \tau_1}(\mb k_1)}\mc P^{(1)}h^\dagger_{ s_2}h^{\phantom\dagger}_{ s_1}\mc P^{(1)}c^\dagger_{\nu_2  \tau_2}(\mb k_2)c^{\phantom\dagger}_{\nu_1  \tau_1}(\mb k_1) \nonumber\\
&&-\frac12 \sum_{\nu_1, \tau_1, s_1} \sum_{\nu_2, \tau_2, s_2} \sum_{\mb k_1,\mb k_2} \sum_L\frac{ t^*_{\nu_2 \tau_2 s_2}(\mb k_2)t_{\nu_1 \tau_1 s_1}(\mb k_1)} {\Delta E_{L}+\varepsilon_{\nu_1 \tau_1}(\mb k_1)}\mc P^{(1)}h^{\phantom\dagger}_{ s_2}\mc P^{(2)}_Lh^\dagger_{ s_1}\mc P^{(1)}c^{\phantom\dagger}_{\nu_2  \tau_2}(\mb k_2)c^\dagger_{\nu_1  \tau_1}(\mb k_1)+\text{h.c.}
\eea
We next compute the required matrix elements between spin-orbital states (where $\bar \sigma=-\sigma$ for $\sigma=\uparrow,\downarrow=+1,-1$),
\begin{eqnarray}\nonumber
\langle s'|h^\dagger_{ s_2}h^{\phantom\dagger}_{ s_1}|s\rangle&=&\delta_{s's_2}\delta_{ss_1}, \quad
\langle s'|h^{\phantom\dagger}_{ s_2}\mc P^{(2)}_{L=0}h^{\dagger}_{ s_1}|s\rangle=\frac13\sigma_2\sigma_1  \delta_{\alpha' \alpha_2}\delta_{\alpha\alpha_1} \delta_{\sigma_2\bar\sigma'}\delta_{\sigma_1\bar\sigma},\\ \nonumber
\langle s'|h^{\phantom\dagger}_{ s_2}\mc P^{(2)}_{L=1}h^{\dagger}_{ s_1}|s\rangle&=&\frac12(\delta_{\alpha_2\alpha_1}\delta_{\alpha'\alpha}-\delta_{\alpha_2\alpha}\delta_{\alpha'\alpha_1})(\delta_{\sigma_2\sigma_1}\delta_{\sigma\sigma'}+\delta_{\sigma_2\sigma}\delta_{\sigma_1\sigma'}),\\ \label{matel}
\langle s'|h^{\phantom\dagger}_{ s_2}\mc P^{(2)}_{L=2}h^{\dagger}_{ s_1}|s\rangle&=&  \delta_{s_2s_1}\delta_{ss'}-\delta_{s_2s}\delta_{s_1s'}- \langle s'|h^{\phantom\dagger}_{s_2} \mc P^{(2)}_{L=0}h_{s_1}^{ \dagger}|s\rangle- \langle s'|h^{\phantom\dagger}_{s_2} \mc P^{(2)}_{L=1}h_{s_1}^{ \dagger}|s\rangle.
\end{eqnarray}
We then obtain the matrix elements of $H_{\rm cot}$ in spin-orbital space as 
\bea
(H_{\rm cot})_{s's}&=&-\frac12 \sum_{\mb k_1\nu_1 \tau_1 } \sum_{\mb  k_2\nu_2 \tau_2 } F_{s's}(\mb k_2,\nu_2,\tau_2;\mb k_1,\nu_1,\tau_1)c^\dagger_{\nu_2  \tau_2}(\mb k_2)c^{\phantom\dagger}_{\nu_1  \tau_1}(\mb k_1) \nonumber\\
&&-\frac12 \sum_{\mb k_1\nu_1 \tau_1 } \sum_{\mb  k_2\nu_2 \tau_2 } \sum_{L=0}^2 G^L_{s's}(\mb k_2,\nu_2,\tau_2;\mb k_1,\nu_1,\tau_1)c^{\phantom\dagger}_{\nu_2  \tau_2}(\mb k_2)c^\dagger_{\nu_1  \tau_1}(\mb k_1)+\text{h.c.}
\eea
with the definitions
\bea \label{auxil}
F_{s's}(\mb k_2,\nu_2,\tau_2;\mb k_1,\nu_1,\tau_1)&=&\frac{ t_{\nu_2 \tau_2 s'}(\mb k_2)t^*_{\nu_1 \tau_1 s}(\mb k_1)}{\Delta E_{0}-\varepsilon_{\nu_1 \tau_1}(\mb k_1)},\\ \nonumber
G^L_{s's}(\mb k_2,\nu_2,\tau_2;\mb k_1,\nu_1,\tau_1)&=&\sum_{s_1,s_2}
\frac{t^*_{\nu_2 \tau_2 s_2}(\mb k_2)t_{\nu_1 \tau_1 s_1}(\mb k_1)}{\Delta E_{L}+\varepsilon_{\nu_1 \tau_1}(\mb k_1)}    \langle s'|h^{\phantom\dagger}_{ s_2}\mc P^{(2)}_{L}h^{\dagger}_{ s_1}|s\rangle.
\eea
\end{widetext}

\begin{figure}[t]
\begin{center}
\includegraphics[width=\columnwidth]{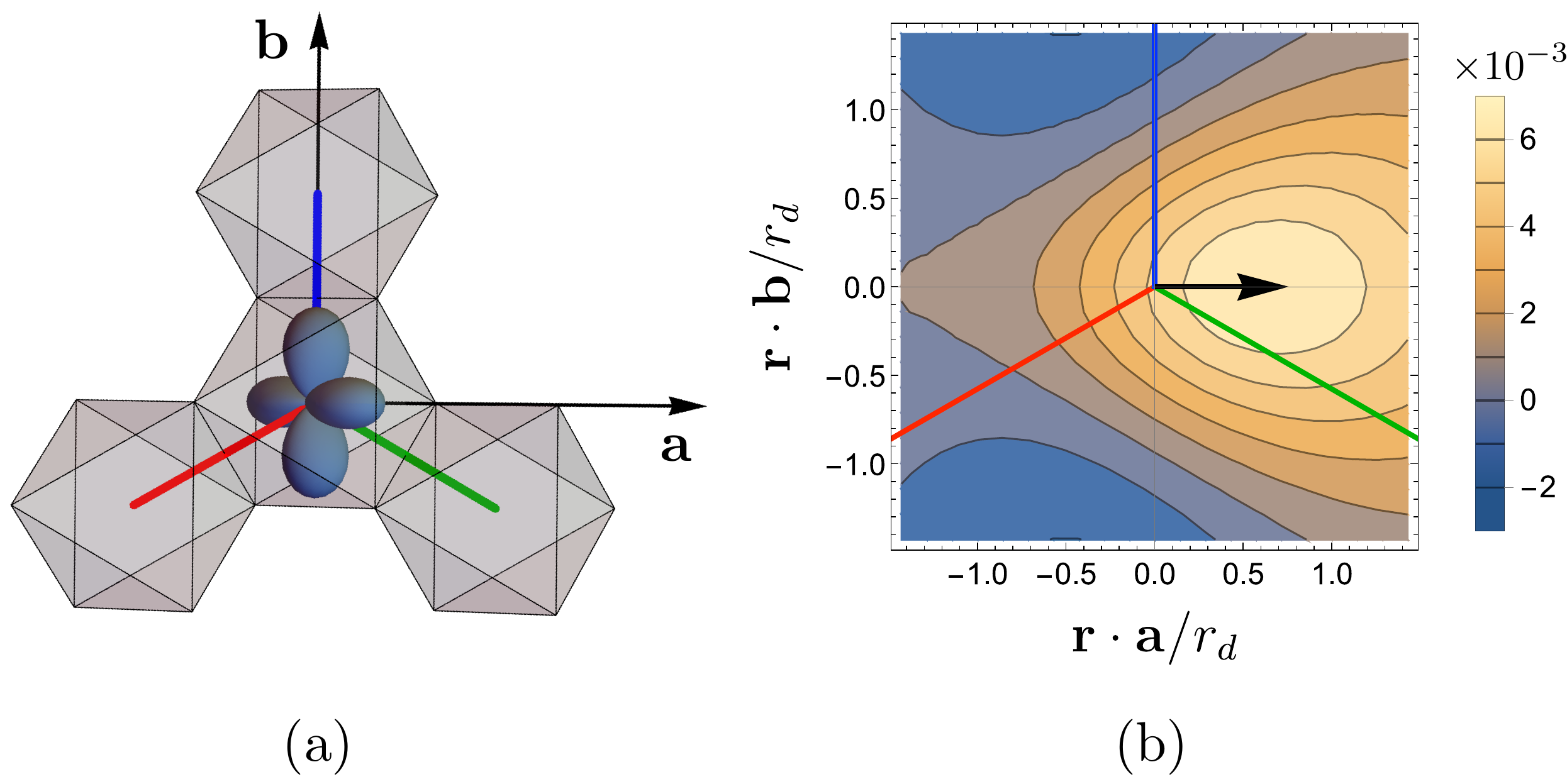}
\end{center}
\caption{Orbital and spatial dependence of tunnel couplings. (a) $xy$-orbital in the edge-sharing octahedra geometry of $\alpha$-RuCl$_3$. The red, green, and blue lines represent the directions of $x$, $y$, and $z$ bonds in the honeycomb plane, respectively. (b) Overlap between the $xy$-orbital at $\mathbf R_j=0$ and the  wave function for electrons in the tip, modeled as an $s$-orbital centered at position $\mathbf r$, see Eq.~(\ref{tiporbital}).  The arrow indicates the point of maximum overlap, corresponding to  the vector $\mathbf v_z$. Here we set $l_s=4l_d$ and $\mb r\cdot \mb c=3l_d$.  }
\label{fig7}
\end{figure}

In a low-energy approach, we can now assume low energies, 
$|\varepsilon_{\nu\tau}(\mb k)|\ll E_g$, for all conduction electron states involved in virtual processes.
For simplicity, we also consider effectively $\mb k$-independent, spin-conserving and spin-independent tunneling amplitudes, 
\be \label{realtunel}
t_{\nu\tau s}(\mb k)=t_{\nu \alpha} \delta_{\tau \sigma},
\ee
with $s=(\alpha,-\sigma)$. 
Tunneling between the substrate ($\nu=B$) and the magnetic layer is modeled by a 
featureless isotropic coupling, $t_{B\alpha}=t_B$.  However, the tunnel couplings connecting the tip ($\nu=A$) to a magnetic site 
depend on the $t_{2g}$-orbital  ($\alpha$) as well as on the relative position between tip and site. For definiteness, we model the $t_{2g}$-orbitals by real wave functions with the proper symmetry. For instance, for the $xy$-orbital centered at $\mb R_j=0$, we take $\Phi_{xy}(\mb r')\propto x'y'e^{-|\mb r'|/l_d}$, where $l_d$ sets the size of the orbital. Here the components of $\mb r'$ refer to the axes fixed by the octahedral environment of the magnetic ion, see  
Fig.~\ref{fig7}(a). In these coordinates, the unit vectors for the conventional crystallographic directions are given by
\be\label{axes}
\mb a=\frac1{\sqrt6}\left(\begin{array}{c} 1\\ 1\\-2\end{array}\right),\quad 
\mb b=\frac1{\sqrt2}\left(\begin{array}{c}-1\\1\\0\end{array}\right),\quad 
\mb c=\frac1{\sqrt3}\left(\begin{array}{c}1\\1\\ 1\end{array}\right),
\ee
where $\mb c$ is perpendicular to the honeycomb plane. As the wave function for the tip at position $\mb r$, we consider \be
\Phi_s(\mb r')\propto e^{ -|\mb r'-\mb r|/l_s},\label{tiporbital}
\ee
with characteristic length $l_s$.

In Fig. \ref{fig7}(b) we show the overlap between $\Phi_{xy}$ and $\Phi_s$ as a function of the tip position, keeping the tip height $\mb r\cdot \mb c>0$ constant and varying the coordinates parallel to the honeycomb plane. The coordinates are scaled by the effective radius of the $t_{2g}$-orbitals, $r_d=\int d^3r'\,r'|\Phi_{\alpha}(\mb r')|^2=7l_d/2$. We denote by $\mb v_\alpha$ the in-plane vector that corresponds to the relative position of maximum overlap between the tip and the $\alpha$-orbital. Note that $\mb v_\alpha$ lies in the direction perpendicular to the $\alpha$-bond. This shift in the position of maximum overlap  can be interpreted in terms of the direction in which the $\alpha$-orbital points above the plane, see Fig.~\ref{fig7}(a). Comparing the ionic radius of Ru$^{3+}$ with the lattice spacing of $\alpha$-RuCl$_3$, we estimate $|\mb v_\alpha|\approx 0.1 a_0$. To capture the orbital and position dependence in the tunnel couplings within a simple analytical expression, we parametrize  $t_{A\alpha}(\mb r,\mb R_j)$  as given in Eq.~\eqref{taj}, with  tunneling length $l_0\sim l_s\lesssim a_0$.

For given $\mb r$ and $\mb R_j$, it is convenient to express the tunnel couplings $t_{A\alpha}$ in terms of spherical
angles $\varphi\in[0,2\pi)$ and $\theta\in[-\pi,\pi]$,
\be
\left(\begin{array}{c} t_{Ax} \\ t_{Ay} \\ t_{Az} \end{array}\right)=
t_A\left(\begin{array}{c} \cos\varphi\sin\theta \\ \sin\varphi\sin\theta \\ \cos\theta \end{array}\right).
\ee
Inserting the above expressions into Eq.~\eqref{auxil} and using Eq.~\eqref{matel}, we finally perform the projection to the $j_{\rm eff}=1/2$ subspace selected by the spin-orbit coupling.  
The corresponding basis states are \cite{Pereira2020}
\begin{eqnarray}\nonumber
|+\rangle &=& \frac{1}{\sqrt{3}}(-|z,\uparrow\rangle-i|y,\downarrow\rangle-|x,\downarrow\rangle),\\ 
|-\rangle &=& \frac{1}{\sqrt{3}}(|z,\downarrow\rangle+i|y,\uparrow\rangle-|x,\uparrow\rangle).\label{basisstates}
\end{eqnarray}
The spin operator appearing in the Kitaev model for this site, $\mb S=\frac12 {\boldsymbol \sigma}$, 
acts in the space spanned by Eq.~\eqref{basisstates}. 
The cotunneling Hamiltonian follows as
\begin{widetext}
\be
H_{\rm cot}=-  \sum_{\mb k_1\nu_1  } \sum_{\mb  k_2\nu_2   } \frac{ t_{\nu_1}t_{\nu_2}}{2\Delta E_{0}}c^\dagger_{\nu_2  }(\mb k_2)\left(f_0\mathbbm 1+\mathbf  f\cdot  \boldsymbol \sigma\right)c^{\phantom\dagger}_{\nu_1 }(\mb k_1) -  \sum_{\mb k_1\nu_1 \tau_1 } \sum_{\mb  k_2\nu_2 \tau_2 } \sum_L \frac{ t_{\nu_1}t_{\nu_2}}{2\Delta E_{L}}c^{\phantom\dagger}_{\nu_2  }(\mb k_2)\left(g^L_0\mathbbm 1+\mathbf  g^L\cdot  \boldsymbol \sigma\right)c^{ \dagger}_{\nu_1 }(\mb k_1)+\text{h.c.},\label{Heff}
\ee
with $f_0$ and $\mb f=(f_x,f_y,f_z)$ given by
\be
f_0=\frac{F_{\uparrow\uparrow}+F_{\downarrow\downarrow}}{2},\quad f_x=\frac{F_{\uparrow\downarrow}+F_{\downarrow\uparrow}}{2},\quad
f_y=i\frac{F_{\uparrow\downarrow}-F_{\downarrow\uparrow}}{2},\quad
f_z=\frac{F_{\uparrow\uparrow}-F_{\downarrow\downarrow}}{2},
\ee
and likewise for $g_0^L$ and $\mathbf g^L$. For given $(\sigma,\sigma')$ indices,
the $2\times 2$ matrices $F_{\sigma\sigma'}$ and
$G^L_{\sigma\sigma'}$ act in conduction electron spin space. 
We find
\bea
F_{\uparrow\uparrow}&=& \frac{1}{3}\begin{pmatrix}
\cos\theta&(1+i)\cos\theta\\
e^{-i\varphi}\sin\theta&(1+i)e^{-i\varphi}\sin\theta
\end{pmatrix},\quad 
        F_{\uparrow\downarrow}=\frac{1}{3}\begin{pmatrix}
            (1-i)\cos\theta&-\cos\theta\\
            (1-i) e^{-i\varphi}\sin\theta&- e^{-i\varphi}\sin\theta
        \end{pmatrix},\quad G^0_{\uparrow\uparrow}
        =\frac{F_{\downarrow\downarrow}}{3},\quad G^0_{\uparrow\downarrow}=-\frac{F_{\uparrow\downarrow}}{3}, \nonumber \\
G^1_{\uparrow\uparrow}&=&\frac{1}{6}\begin{pmatrix}
(1-i) e^{i\varphi}\sin\theta+2\cos\theta&(1+i)\cos\theta\\
    e^{-i\varphi}\sin\theta&(\sin\varphi+\cos\varphi)\sin\theta
\end{pmatrix},\quad G^1_{\uparrow\downarrow}=\frac{1}{6}\begin{pmatrix}
            e^{-i\varphi}\sin\theta&(\sin\varphi+\cos\varphi)\sin\theta\\
            (1-i) e^{-i\varphi}\sin\theta&-(1-i)\cos\theta
        \end{pmatrix}, \nonumber \\
G^2_{\uparrow\uparrow}&=&[(\cos\varphi+\sin\varphi)\sin\theta+\cos\theta] \mathbb{1}-F_{\uparrow\uparrow}-G^0_{\uparrow\uparrow}-G^1_{\uparrow\uparrow},\quad G^2_{\uparrow\downarrow}=-F_{\uparrow\downarrow}-G^0_{\uparrow\downarrow}-G^1_{\uparrow\downarrow}.
\eea
The remaining matrices are obtained by using a time-reversal operation,
\begin{equation}
F_{\downarrow\downarrow}=\tau_yF^*_{\uparrow\uparrow}\tau_y,\quad F_{\downarrow\uparrow}=-\tau_yF^*_{\uparrow\downarrow}\tau_y,\quad G^L_{\downarrow\downarrow}=\tau_y\left(G^L\right)^*_{\uparrow\uparrow}\tau_y,\quad G^L_{\downarrow\uparrow}=-\tau_y\left(G^L\right)^*_{\uparrow\downarrow}\tau_y,
\end{equation}
\end{widetext}
with Pauli matrices $\boldsymbol \tau$ in conduction electron spin space.
In the second  term  of Eq.~(\ref{Heff}), we now  use
\[
c^{\phantom\dagger}_{\nu_2 \tau_2 }(\mb k_2)c^{ \dagger}_{\nu_1 \tau_1}(\mb k_1)=-c^{ \dagger}_{\nu_1 \tau_1}(\mb k_1)c^{\phantom\dagger}_{\nu_2 \tau_2 }(\mb k_2)+\delta_{\nu_1\nu_2}\delta_{\tau_1\tau_2}\delta_{\mb k_1\mb k_2}.
\]
The factor  $\delta_{\tau_1\tau_2}$ in the last term implies a trace over  
the $2\times2$ matrices for conduction electrons. As a result, only the identity can contribute.
We thereby obtain the cotunneling Hamiltonian \eqref{Hcot}, 
where $\Psi_A(\mb r)=\sum_{\mb k} c_A(\mb k)$ is a real-space two-component spinor field describing 
conduction electrons on the tip at position $\mb r$. Likewise, $\Psi_B(\mb R)$ refers
to the substrate spinor field below the site with position $\mb R$.
Cotunneling processes are then characterized by the transition matrices $T_0$ and $T^\alpha$, 
with $\mb T=(T^x,T^y,T^z)$, which act in conduction electron spin space and are given by
\bea\nonumber
T_0 &=& -\frac{t_A t_B}{\Delta E_0} f_0 + \sum_{L=0}^2 \frac{t_A t_B}{\Delta E_L} g_0^L,\\ 
\mb T&=&  -\frac{t_A t_B}{\Delta E_0} \mb f + \sum_{L=0}^2 \frac{t_A t_B}{\Delta E_L} \mb g^L.\label{T0}
\eea
All matrix elements  scale $\propto t_A t_B/U$, where individual
contributions carry $\frac{J_H}{U}$-dependent factors.
We emphasize that $T_0$ and $\mb T$ depend on $\mb r-\mb R_j$, with the tip (site) position $\mb r$ ($\mb R_j$).

The above expressions can be simplified considerably when neglecting the orbital-dependent shifts $\mb v_\alpha$ in Eq.~\eqref{taj}.  This approximation becomes exact for a tip placed right on top of a
magnetic site, and otherwise causes quantitative ($\approx 10$\%) deviations in the tunnel couplings.  We then obtain 
\begin{eqnarray*}
f_0&=&\frac{1}{2\sqrt3}\mathbbm 1,\quad f_\alpha=\frac{1}{3\sqrt3}(\tau^x+\tau^y+\tau^z)-\frac{1}{2\sqrt3} \tau^\alpha,\\ \nonumber
g^0_0&=&\frac{1}{6\sqrt3}\mathbbm 1,\quad g^0_\alpha=-\frac{29}{2\sqrt3}(\tau^x+\tau^y+\tau^z)+\frac{1}{6\sqrt3}\tau^\alpha,\\ \nonumber
g^1_0&=&\frac{1}{2\sqrt3}\mathbbm 1,\quad g^1_\alpha=\frac{13}{2\sqrt3}\tau^\alpha,\\ \nonumber
g^2_0&=&\frac{23}{3\sqrt3}\mathbbm 1,\quad g^2_\alpha=-\frac{55}{9\sqrt3}(\tau^x+\tau^y+\tau^z)+\frac{26}{3\sqrt3}\tau^\alpha,
\end{eqnarray*}
and $H_{\rm cot}$ takes the form \eqref{cotfin}, 
where we define the $\frac{J_H}{U}$-dependent coefficients ($j=0,1,2$) 
\be\label{cotrate}
 \eta_{j}= \frac{U}{2\sqrt3 \Delta E_0} \zeta_j +\sum_{L=0}^2 \frac{U}{2\sqrt3 \Delta E_L} \zeta_j^L
\ee
with $\Delta E_L$  in Eq.~\eqref{DeltaEL} and the numbers
\begin{eqnarray*}
\zeta_0 &=& 1 , \quad \zeta_0^1 = \frac13 ,\quad \zeta_0^2 =1, \quad \zeta_0^3= \frac{46}3,\\
\nonumber
\zeta_1&=&-\frac12,\quad \zeta_2=\frac{1}{3},\quad \zeta^0_1=\frac{1}{6},\quad \zeta^0_2=-\frac{1}{9},\quad \zeta^1_1=0,\\
\zeta^1_2&=&\frac{1}{6},\quad \zeta^2_1=\frac{1}{3},\quad \zeta^2_2=-\frac{7}{18}.
\end{eqnarray*}


\begin{thebibliography}{99}
\bibitem{footnote}  Some of the relations in Sec.~\ref{sec3}, in particular, Eqs.~\eqref{Thouless} and \eqref{eq:dsf_matrixelements}, assume that $|0_a\rangle$ and $|0_b\rangle$ are the respective vacuum states without extra excitations to satisfy the fermion parity constraint.  The corresponding expressions in the presence 
of such excitations follow accordingly.
\end{thebibliography}
\end{document}